\documentclass[12pt]{article}

\usepackage{amsfonts,amsmath,amssymb}
\usepackage{hyperref}
\usepackage{cite}
\usepackage{epsfig}
\usepackage{paralist}
\usepackage{fancyhdr}
\usepackage{tikz}
\usepackage{tkz-euclide}
\usetikzlibrary{decorations.pathmorphing,patterns,calc,snakes,arrows}

\usepackage{graphicx}
\usepackage{xcolor}
\numberwithin{equation}{section}
\usepackage[vcentermath]{youngtab}
\usepackage{etex}
\usepackage{braket}
\usepackage{float}

\def\spa#1{\phantom{\fbox{\rule[-#1cm]{0cm}{0cm}}}}

\def\be{\begin{equation}}
\def\ee{\end{equation}}
\def\bea{\begin{eqnarray}}
\def\eea{\end{eqnarray}}

\def\del{\partial}

\renewcommand{\thefootnote}{\fnsymbol{footnote}}

\def\TTbar{\mbox{$T\bar{T}$}}

\makeatletter
\g@addto@macro\bfseries{\boldmath}
\makeatother

\def\p{\partial}
\def\pb{\bar{\partial}}

\def\xb{{\bar{x}}}

\def\zb{{\bar{z}}}

\def\Tb{{\bar{T}}}
\def\Xb{{\bar{X}}}
\def\Zb{{\bar{Z}}}

\def\Im{\mathop{\mathrm{Im}}\nolimits}

\def\Re{\mathop{\mathrm{Re}}\nolimits}

\begin{document}

\hfuzz=100pt
\title{{\Large \bf{Conformal field theory on $T\Tb$-deformed space\\
 and \\
 correlators from dynamical coordinate transformations}}}
\date{}
\author{Shinji Hirano$^{a, c}$\footnote{
	e-mail:
	\href{mailto:shinji.hirano@wits.ac.za}{shinji.hirano@gmail.com}} ~and
 Masaki Shigemori$^{b, c}$\footnote{
	e-mail:
	\href{mailto:masaki.shigemori@nagoya-u.jp}{masaki.shigemori@nagoya-u.jp}}}
\date{}

\maketitle

\thispagestyle{fancy}
\rhead{YITP-24-17}
\cfoot{}
\renewcommand{\headrulewidth}{0.0pt}

\vspace*{-1cm}
\begin{center}
$^{a}${{\it School of Science, Huzhou University}}
\\ {{\it Huzhou 313000, Zhejiang, China}}
  \spa{0.5} \\
$^b${{\it Department of Physics, Nagoya University}}
\\ {{\it Furo-cho, Chikusa-ku, Nagoya 464-8602, Japan}}
\spa{0.5}  \\
$^c${{\it Center for Gravitational Physics and Quantum Information (CGPQI)}}
\\ {{\it  Yukawa Institute for Theoretical Physics, Kyoto University}}
\\ {{\it Kitashirakawa-Oiwakecho, Sakyo-ku, Kyoto 606-8502, Japan}}
\spa{0.5}  

\end{center}

\begin{abstract}

We study the map between two descriptions of the $T\Tb$ deformation of conformal field theory (CFT): One is the defining description
as a deformation of CFT by the $T\Tb$-operator. The other is an alternative description as the undeformed CFT on the dynamical $T\Tb$-deformed space
that backreacts to the state or operator insertions, reminiscent of the theory of gravity. 
Instead of adopting the topological gravity description, we develop a more literal CFT-based operator formalism that facilitates systematic and straightforward computations of the $T\Tb$-deformation of the stress tensor, operators, and their correlators, while rederiving known results in the literature.
Along the way, we discuss the backreaction to the $T\Tb$-deformed space in response to local operators and exhibit the hard-disk and free-space structures in the UV-cutoff and Hagedorn phases, respectively, suggested by Cardy-Doyon and Jiang. 
To capitalize on the alternative description of the $T\Tb$-deformed CFT, we focus on the correlators of semi-heavy operators, i.e., the operators of large conformal dimension $\Delta\gg\sqrt{c}$, and show an intuitive and simple way to obtain the $T\Tb$-deformed correlators from those of the undeformed CFT on the $T\Tb$-deformed space via dynamical coordinate transformations.
This may have implications in the holographic dual description,  pointing towards a working dictionary for a class of matter correlators in the cutoff AdS picture.

\end{abstract}

\renewcommand{\thefootnote}{\arabic{footnote}}
\setcounter{footnote}{0}

\newpage

\tableofcontents


\section{Introduction}
\label{Sec:Introduction}

As a quantum field theory (QFT), the difficulty of quantizing General Relativity (GR) is rooted in the fact that the gravitational constant $G_N$ has the (classical) dimension, length-squared, in four spacetime dimensions. Namely, the gravitational coupling is irrelevant in the sense of renormalization group (RG) and keeps becoming stronger as the energy scale is increased. This implies uncontrollable divergences at short distance scales unless it is somehow rendered asymptotically safe \cite{Weinberg:1978,Weinberg:1980gg} or UV-completed by new degrees freedom, such as strings, that effectively regulate short distance divergences.

Recently, there emerged a class of two-dimensional QFTs with irrelevant couplings, or interactions, which appear to be UV-complete, the prototype of which is known as the $T\Tb$ deformation \cite{Zamolodchikov:2004ce,Smirnov:2016lqw, Cavaglia:2016oda}. As this name suggests, one typically views these interactions as deformations of renormalizable QFTs, and these deformations are known to preserve integrability when the undeformed QFTs are integrable. 
To understand the most basic properties of these theories, we focus on the $T\Tb$ deformation among other deformations which, as the notation indicates, is a bilinear of the stress tensor components and can be added to any local QFTs in a model-independent way. To be as universal as possible, we further restrict our attention to the case of the $T\Tb$-deformed conformal field theory (CFT) whose only scale enters via the $T\Tb$-coupling $\mu$ of dimension length-squared. 
As one might expect, the $T\Tb$-deformed CFT shows signs of non-locality at short distance scales associated with the $T\Tb$-coupling $\mu$. For example, as we will see in Section \ref{sec:TTbardeformedspace}, there is a mechanism by which the short distance is cut off and that is presumably related to the fact that the UV divergences in this power-counting nonrenormalizable theory are rendered under control. In this regard, it is hoped that the nature of the $T\Tb$-deformed theory has some bearings on the short-distance physics which quantum gravity might share.

\begin{figure}[h!]
\centering\includegraphics[height=1.0in]{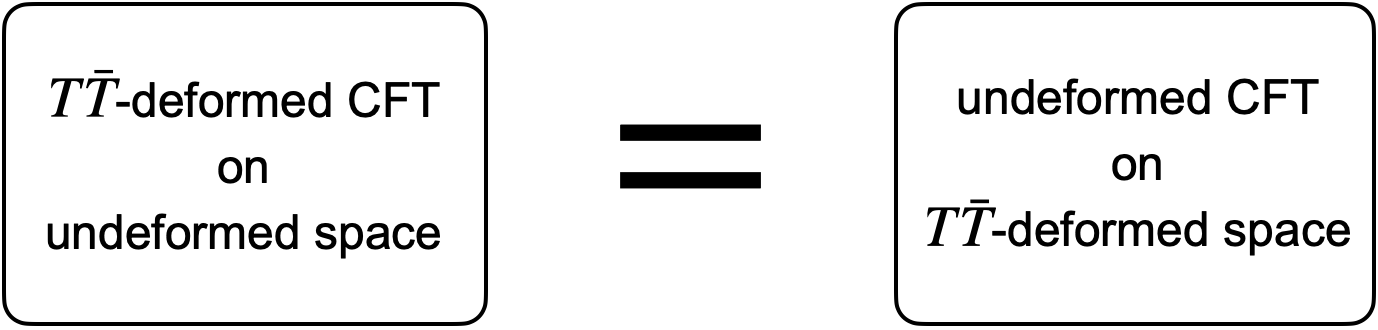}
\caption{\sl Two descriptions of the $T\Tb$-deformation of CFT: It can be viewed either as a deformation of the theory or a deformation of the space. The $T\Tb$-deformed space is dynamical in the sense that it backreacts to the state or operator insertions. It has an elegant description in terms of the flat space JT gravity \cite{Dubovsky:2012wk, Dubovsky:2017cnj,Dubovsky:2018bmo}, or equivalently, a ghost-free massive gravity \cite{Tolley:2019nmm}. In this paper, however, we develop a less sophisticated, yet complementary, operator formalism based on a literal interpretation of this equivalence. To be more precise and not to mislead, we note that the two descriptions are only the endpoints of an infinite number of in-between hybrid descriptions as discussed in Section \ref{sec:map},.}
\label{fig:TwoDescript}
\end{figure}

Our main interest in this paper is in the following observation and we explore several aspects of it:
There are two descriptions of the $T\Tb$ deformation of CFT as illustrated in Figure \ref{fig:TwoDescript}.
One is the defining description as a deformation of CFT by the $T\Tb$-operator \cite{Zamolodchikov:2004ce}. The other is an alternative description as the \emph{un}deformed CFT on the $T\Tb$-deformed space that backreacts to the state or operator insertions \cite{Conti:2018tca, Cardy:2019qao}.
Our main objective is to study the map between the two more in detail than in the earlier works and to better understand the latter unconventional description, {\it i.e.}, a QFT living on a \emph{dynamical} space, reminiscent of the theory of (quantum) gravity.
In fact, it was shown in \cite{Dubovsky:2012wk, Dubovsky:2017cnj,Dubovsky:2018bmo} that the $T\Tb$-deformed CFT can be described by the undeformed CFT coupled to the Jackiw-Teitelboim (JT) gravity \cite{Jackiw:1984je,Teitelboim:1983ux} in the flat space limit,  or equivalently, a ghost-free massive gravity \cite{Tolley:2019nmm}. In this work, however, we interpret this equivalence more literally and develop a CFT-based operator formalism which we believe provides a complementary view to the JT gravity description.

In Section \ref{sec:map}, we first review the basic ideas of \cite{Conti:2018tca, Cardy:2019qao} on the different descriptions of the $T\Tb$ deformation and give our own account of how they are mapped to each other.  This introduces a notion of dynamical coordinate transformation and the $T\Tb$-deformed space. In particular, from the maps or dynamical coordinate transformations, we give a simple rederivation of the recursion relations for the $T\Tb$-deformed stress tensor previously found in the literature. These set the basis for the analyses and discussions in the subsequent sections. 
We then begin to study the physical and technical implications inferred from the dynamical coordinate transformations. 
In Section \ref{sec:TTbardeformedspace}, we discuss the short distance structure of the $T\Tb$-deformed space due to the backreaction  in response to local operators and exhibit the peculiar properties of the $T\Tb$-deformed space suggested earlier in \cite{Cardy:2020olv, Jiang:2020nnb}.
In Sections~\ref{sec:correlators} and~\ref{sec:heavycorrelators}, we study the $T\Tb$-deformed correlators, developing the operator formalism of CFT on the $T\Tb$-deformed space.
In particular, we show in Section \ref{sec:heavycorrelators} that, in a semiclassical limit, the $T\Tb$-deformed correlators can be computed, in an intuitive and simple way, from those of CFT on the $T\Tb$-deformed space via dynamical coordinate transformations.
In Section \ref{sec:2ptcutoffAdS}, we discuss the implications of the results in Section \ref{sec:heavycorrelators} in the holographic description of the $T\Tb$-deformed CFT\@.
In Section \ref{sec:Discussions}, we give a brief summary of the results and add a further discussion on the CFT on the $T\Tb$-deformed space.

\section{The map between two descriptions}
\label{sec:map}

Let us denote the $T\Tb$-deformed CFT on $\mathbb{R}^2$ by ${\cal T}^{(\mu)}[\mathbb{R}^2]$. 
The $T\Tb$-deformation is defined by the infinitesimal deformation of the action
\begin{align}\label{infinitesimal_deformed_action}
S[\mu+\delta\mu]=S[\mu]+{\delta\mu\over\pi^2}\int_{\mathbb{R}^2}d^2x{\cal O}^{(\mu)}_{T\Tb}
\quad\mbox{where}\quad {\cal O}^{(\mu)}_{T\Tb}=T^{(\mu)}\Tb^{(\mu)}-(\Theta^{(\mu)})^2=-\det T^{(\mu)}_{ij},
\end{align}
where $T^{(\mu)}=T_{zz}^{(\mu)}$, $\Tb^{(\mu)}=T_{\zb\zb}^{(\mu)}$, and $\Theta^{(\mu)}=T_{z\zb}^{(\mu)}$ for the flat metric $ds^2=dzd\zb$. So the $T\Tb$-deformed CFT, ${\cal T}^{(\mu)}[\mathbb{R}^2]$, on $\mathbb{R}^2$ is obtained by the iteration of the infinitesimal transformations starting with the undeformed CFT, ${\cal T}^{(0)}[\mathbb{R}^2]$. 
Note that at every step of the infinitesimal deformation, the stress tensor gets deformed and  must be updated in the next step. This is the reason why the $T\Tb$-deformation is defined in the infinitesimal form. We will develop an algorithm to systematically compute the explicit form of the deformed stress tensor to an arbitrary perturbative order in the $T\Tb$ coupling $\mu$.

As stated in the introduction, the $T\Tb$-deformed CFT on $\mathbb{R}^2$
is equivalent to the undeformed CFT on the $T\Tb$-deformed
$\mathbb{R}^2$, denoted by $\mathbb{R}_{(0|\mu)}^2$, equipped with the
operator-valued dynamical coordinates, $Z^{(\mu)}$ and
$\bar{Z}^{(\mu)}$, which are functions of $(z,\zb)$.  This equivalence
between the two descriptions can be expressed as ${\cal
T}^{(\mu)}[\mathbb{R}_{(\mu|0)}^2]={\cal
T}^{(0)}[\mathbb{R}_{(0|\mu)}^2]$, where $\mathbb{R}_{(\mu|0)}^2$ means
that we are deforming by $\mu$ the theory defined on the undeformed $\mathbb{R}^2$.
More generally, we can consider deforming by $\mu_1$ the theory defined
on the space that is deformed by $\mu_2$, so that the net
$T\Tb$-deformation coupling is $\mu=\mu_1+\mu_2$. In this case,
the space is denoted by 
$\mathbb{R}_{(\mu_1|\mu_2)}^2$ and the theory on by
${\cal
T}^{(\mu_1)}[\mathbb{R}_{(\mu_1|\mu_2)}^2]$.\footnote{This notation is slightly
redundant in the sense that the index of ${\cal T}^{(\mu_1)}$ is always
the same as the first index of $\mathbb{R}_{(\mu_1|\mu_2)}^2$. } In this more general setting, there are an infinite number of
equivalences, ${\cal
T}^{(\lambda_1)}[\mathbb{R}_{(\lambda_1|\lambda_2)}^2]={\cal
T}^{(\mu_1)}[\mathbb{R}_{(\mu_1|\mu_2)}^2]$ with
$\lambda_1+\lambda_2=\mu_1+\mu_2$. In subsection~\ref{sec:infinitesimal}, we exploit the equivalence, ${\cal
T}^{(\mu)}[\mathbb{R}_{(\mu|\delta\mu)}^2]={\cal
T}^{(\mu+\delta\mu)}[\mathbb{R}_{(\mu+\delta\mu|0)}^2]$, to derive the
recursion relations for the deformed stress tensor from which we can
systematically find the explicit form of the deformed stress tensor in
the perturbative $\mu$-expansion.  In subsection~\ref{sec:finite}, we
focus on the equivalence ${\cal T}^{(\mu)}[\mathbb{R}_{(\mu|0)}^2]={\cal
T}^{(0)}[\mathbb{R}_{(0|\mu)}^2]$ to study the undeformed CFT on the
deformed space.

\subsection{The infinitesimal map}
\label{sec:infinitesimal}

We first discuss the infinitesimal map between the undeformed ordinary
coordinate $z\equiv Z^{(\mu|0)}$ and deformed dynamical coordinate $Z^{(\mu|\delta\mu)}$,
and its implication on the deformed stress tensor components
$T^{(\mu)}$, $\Tb^{(\mu)}$, and $\Theta^{(\mu)}$.  The infinitesimal
deformation takes the ${\cal T}^{(\mu)}[\mathbb{R}^2]$ theory to ${\cal
T}^{(\mu+\delta\mu)}[\mathbb{R}^2]={\cal
T}^{(\mu+\delta\mu)}[\mathbb{R}_{(\mu+\delta\mu|0)}^2]$ which can be
mapped to ${\cal T}^{(\mu)}[\mathbb{R}_{(\mu|\delta\mu)}^2]$. In the
latter description, the dynamical coordinate of
$\mathbb{R}_{(\mu|\delta\mu)}^2$ is given by
\cite{Cardy:2019qao}\footnote{Our convention is that $d^2x=2d(\Re
x)\wedge d(\Im x)={i}dx\wedge d\bar{x}$ and $\delta^2(x)={1\over
2}\delta(\Re x)\delta(\Im x)$.}
\begin{align}\label{infinitesimalZmap}
z\quad\longmapsto\quad Z^{(\mu|\delta\mu)}(z,\zb)=z+{\delta\mu\over 2\pi^2}\int_{\mathbb{R}^2}d^2x{\bar{T}^{(\mu)}(x,\bar{x})\over z-x}+{\cal O}(\delta\mu^2)\ ,
\end{align}
where   we can alternatively write $z$ as $z=Z^{(\mu|0)}$.
This expression is equivalent to Cardy's line integral representation in \cite{Cardy:2019qao}, as will be elaborated in Appendix \ref{app:lineintegral}, and can be derived from the random geometry description
\cite{Cardy:2018sdv} as follows: By using the Hubbard-Stratonovich
transformation, the $T\Tb$-deformation
\eqref{infinitesimal_deformed_action} can be expressed as
\begin{align}\label{HS}
e^{-S[\mu+\delta\mu]}&=e^{-S[\mu]}\int [dh]e^{-{1\over 8\delta\mu}\int d^2x \epsilon^{ik}\epsilon^{jl}h_{ij}h_{kl}-{1\over 4\pi}\int d^2x h_{ij}T^{(\mu)ij}}
\end{align}
where $[dh]$ is the properly normalized integration measure for the auxiliary symmetric tensor $h_{ij}$. For the infinitesimal transformation, $\delta\mu\ll \mu$ and the $h$-integrations are dominated by the saddle point
\begin{align}\label{saddleeqn}
h^{\ast}_{ij}=-{\delta\mu\over\pi}\epsilon_{ik}\epsilon_{jl}T^{(\mu)kl}\ .
\end{align} 
This implies that $h^{\ast}_{ij}=\partial_i\alpha_j+\partial_j\alpha_i $ with
\begin{align}\label{saddle}
\alpha^z=2\alpha_{\zb}={\delta\mu\over 2\pi^2}\int_{\mathbb{R}^2}d^2x{\bar{T}^{(\mu)}(x,\bar{x})\over z-x}
\end{align}
where $z=x+iy$ and $\bar{\partial}{1\over z-x}=2\pi\delta^2(z-x)$. We note that 
\begin{align}\label{ThetabyT}
\Theta^{(\mu)}={1\over 2\pi}\int_{\mathbb{R}^2}d^2x{\bar{T}^{(\mu)}(x,\bar{x})\over (z-x)^2}={1\over 2\pi}\int_{\mathbb{R}^2}d^2x{T^{(\mu)}(x,\bar{x})\over (\zb-\bar{x})^2} 
\end{align}
by using integration by parts and the conservation law $\bar{\partial}T^{(\mu)}+\partial\Theta^{(\mu)}=\partial\Tb^{(\mu)}+\bar{\partial}\Theta^{(\mu)}=0$. In other words,  the saddle point solution \eqref{saddle} respects the conservation law. With the form of the $T\Tb$-deformed action \eqref{HS} and the definition of the stress tensor $\sqrt{g}T_{ij}=-2\delta S/\delta g^{ij}$, the $T\Tb$-deformation can then be interpreted as the (dynamical) coordinate transformation
\begin{align}
(z, \zb) \quad\longmapsto \quad (z + \alpha^z, \zb+\alpha^{\zb})\ ,
\end{align}
yielding the infinitesimal transformation law \eqref{infinitesimalZmap}.

\subsubsection{The stress tensor flow and deformation}
\label{sec:stressdef}

As an application of the dynamical coordinate transformation \eqref{infinitesimalZmap}, we provide a somewhat simpler derivation of the stress tensor deformation discussed in \cite{Cardy:2019qao}. For this purpose, we consider the equivalence, ${\cal T}^{(\mu+\delta\mu)}[\mathbb{R}_{(\mu+\delta\mu|0)}^2]={\cal T}^{(\mu)}[\mathbb{R}_{(\mu|\delta\mu)}^2]$.
The conservation of the stress tensor in these two descriptions implies that
\begin{align}\label{conservation_inf}
\underbrace{\p_a T^{(\mu+\delta\mu)}\mbox{}^a_{\mbox{  }b}(x)}_{\text{deformed $T$ on undeformed space}}
=\underbrace{\tilde{\p}_a T^{(\mu)}\mbox{}^a_{\mbox{  }b}(\tilde{x})}_{\text{``undeformed'' $T$ on deformed space}}
=0\ ,
\end{align} 
where we introduced a notation $(x^1, x^2)=(z, \zb)$ and $(\tilde{x}^1, \tilde{x}^2)=(Z^{(\mu|\delta\mu)},\bar{Z}^{(\mu|\delta\mu)})$. Using 
\begin{align}\label{zZJacobian}
\left(
\begin{array}{c}
dz\\
d\bar{z}
\end{array}
\right)
=\left(
\begin{array}{cc}
1+{\delta\mu\over \pi}\Theta^{(\mu)}(z,\zb) & -{\delta\mu\over \pi}\Tb^{(\mu)}(z,\zb) \\
-{\delta\mu\over \pi}T^{(\mu)}(z,\zb) & 1+{\delta\mu\over \pi}\Theta^{(\mu)}(z,\zb)
\end{array}
\right)
\left(
\begin{array}{c}
dZ^{(\mu|\delta\mu)}\\
d\bar{Z}^{(\mu|\delta\mu)}
\end{array}
\right)+{\cal O}(\delta\mu^2)\ ,
\end{align}
one can find that
\begin{equation}
\begin{aligned}
0&=\pb T^{(\mu+\delta\mu)}+\p \Theta^{(\mu+\delta\mu)}\\
&=\pb\left(T^{(\mu)}(\tilde{x})-{2\delta\mu\over\pi}\Theta^{(\mu)}T^{(\mu)}\right)
+\p\left(\Theta^{(\mu)}(\tilde{x})-{\delta\mu\over\pi}\left(\Tb^{(\mu)}T^{(\mu)}+(\Theta^{(\mu)})^2\right)\right)\ ,
\end{aligned}
\end{equation}
where the stress tensor components are the functions of $x$ unless otherwise indicated. This then reads the flow equations for the stress tensor components:
\begin{align}
T^{(\mu+\delta\mu)}(x)&=T^{(\mu)}(\tilde{x})-{2\delta\mu\over\pi}\Theta^{(\mu)}T^{(\mu)}(x)+{\cal O}(\delta\mu^2)\ ,\label{Tinf_flow}\\
\Theta^{(\mu+\delta\mu)}(x)&=\Theta^{(\mu)}(\tilde{x})-{\delta\mu\over\pi}\left(\Tb^{(\mu)}T^{(\mu)}(x)+(\Theta^{(\mu)}(x))^2\right)+{\cal O}(\delta\mu^2)\ .\label{Thetainf_flow}
\end{align}
From these equations, using further the infinitesimal transformation law \eqref{infinitesimalZmap} as well as \eqref{ThetabyT}, we find the recursion relation for the deformed stress tensor components: 
\begin{equation}
\begin{aligned}\label{DeformedTformula}
T^{(\mu)}(z,\zb)=&\, T(z)
+ \int^{\mu}_0{d\lambda\over 2\pi^2}\Biggl[
2\int_{\mathbb{R}^2} d^2x{\p\Tb^{(\lambda)}(x,\bar{x})T^{(\lambda)}(z, \bar{z})\over z-x}\\
&+\int_{\mathbb{R}^2}d^2x {\Tb^{(\lambda)}(x, \bar{x})\p T^{(\lambda)}(z,\zb)\over z-x}
+\int_{\mathbb{R}^2}d^2x {T^{(\lambda)}(x, \bar{x})\pb T^{(\lambda)}(z,\zb)\over \zb-\bar{x}}\Biggr]\ ,
\end{aligned}
\end{equation}
where $T(z)=T^{(0)}(z,\zb)$ is the holomorphic stress tensor of the undeformed CFT, and
\begin{equation}
\begin{aligned}\label{DeformedThetaformula}
\Theta^{(\mu)}(z,\zb)&= \int^{\mu}_0{d\lambda\over 2\pi^2} \Biggl[
 -2\pi\left({\Tb^{(\lambda)}(z, \bar{z})T^{(\lambda)}(z,\bar{z})}-\Theta^{(\lambda)}(z,\bar{z})^2\right)\\
&+\partial_z\int_{\mathbb{R}^2}d^2x{\Tb^{(\lambda)}(x, \bar{x})\Theta^{(\lambda)}(z,\bar{z})\over z-x}
+\partial_{\zb}\int_{\mathbb{R}^2}d^2x{T^{(\lambda)}(x, \bar{x})\Theta^{(\lambda)}(z,\bar{z})\over \bar{z}-\bar{x}}\Biggr]\ ,
\end{aligned}
\end{equation}
where $\Theta^{(0)}(z,\zb)=0$ since the CFT stress tensor is traceless.

A few remarks are in order: there are various ways to express the recursion relations. The formula \eqref{DeformedTformula} is particularly convenient for the systematic computation of the deformed stress tensor order by order in the $T\Tb$ coupling $\mu$. As an illustration, the explicit expressions for the stress tensor deformation, to second order, are provided in Appendix \ref{app:secondorder}\@. It should be noted that the composite operators that appear in the formula need to be regularized and we adopt the point-splitting regularization and renormalization.  
The formula \eqref{DeformedThetaformula} for the trace makes it easier to see its equivalence to the more familiar form of the flow equation \cite{Kraus:2018xrn}
\begin{align}\label{floweqn}
\Theta^{(\mu)}(z,\zb)=-{\mu\over\pi}\left[T^{(\mu)}\Tb^{(\mu)}(z,\zb)-\Theta^{(\mu)}(z,\zb)^2\right]={\mu\over\pi}\det T^{(\mu)}_{ij}\ .
\end{align}
We will discuss more on this point in Section \ref{sec:finite}. 
As an alternative to \eqref{DeformedTformula}, for example, there exists a more concise expression
\begin{align}\label{DeformedTconcise}
T^{(\mu)}(z,\zb)=T(z)-i\p J^{(\mu)}_z(z,\zb).
\end{align}
where $J^{(\mu)}_z$ is a spin 1 current
\begin{align}
J^{(\mu)}_z(z,\zb)=i \int^{\mu}_0{d\lambda\over 2\pi^2}\left[\int_{\mathbb{R}^2} {d^2x\Tb^{(\lambda)}(x, \bar{x})T^{(\lambda)}(z,\zb)\over z-x}
-\int_{\mathbb{R}^2} {d^2xT^{(\lambda)}(x, \bar{x})\Theta^{(\lambda)}(z,\zb)\over \zb-\bar{x}}\right]
\end{align}
which turns out to be conserved, $\pb J^{(\mu)}_z+\p J^{(\mu)}_{\zb}=0$ with $J^{(\mu)}_{\zb}=\bar{J}^{(\mu)}_z$. In terms of this current, the trace is given by
\begin{align}
\Theta^{(\mu)}(z,\zb)=i\pb J^{(\mu)}_{z}=-i\p J^{(\mu)}_{\zb}\ .
\end{align}
Note that with the flow equation \eqref{floweqn}, this shows that the $T\Tb$-deformation is a total derivative and thus topological in some sense as observed in \cite{Cardy:2019qao, Cardy:2018sdv}.
In this paper, instead of exploring aspects of this tantalizing property, we focus more on the practical use of these expressions.

To make a point of the formulas such as \eqref{DeformedTformula} and \eqref{ThetabyT}, they provide a new technique for a systematic and straightforward computation of the stress tensor correlators to an arbitrary order, going beyond the low-order analyses in \cite{Kraus:2018xrn, Aharony:2018vux, Hirano:2020ppu}.

\subsection{The finite map}
\label{sec:finite}

The infinitesimal map \eqref{infinitesimalZmap} in the previous section is of good practical use since it yielded the recursion relations for the stress tensor components that allow us to perform a systematic and straightforward computation of the $T\Tb$-deformed stress tensor in terms of the undeformed CFT (anti-)holomorphic stress tensor components $T(z)$ and $\Tb(\zb)$. In this section, we discuss the direct finite map between the $T\Tb$-deformed CFT on the undeformed space, ${\cal T}^{(\mu)}[\mathbb{R}^2_{(\mu|0)}]={\cal T}^{(0)}[\mathbb{R}^2]$, and the undeformed CFT on the $T\Tb$-deformed space, ${\cal T}^{(0)}[\mathbb{R}_{(0|\mu)}^2]$.

The dynamical coordinate of $\mathbb{R}_{(0|\mu)}^2$ is given by \cite{Dubovsky:2012wk, Dubovsky:2017cnj,Dubovsky:2018bmo, Conti:2018tca}
\begin{align}\label{finiteZmap}
z\quad\longmapsto\quad Z^{(\mu)}(z,\zb)=z+{\mu\over 2\pi^2}\int_{\mathbb{R}^2}d^2x{\bar{T}^{(\mu)}(x,\bar{x})\over z-x}\ ,
\end{align}
where we used a simplified notation $Z^{(\mu)}\equiv Z^{(0|\mu)}$ and $z=Z^{(\mu|0)}$.
This is equivalent to the following infinitesimal transformation from ${\cal T}^{(0)}[\mathbb{R}_{(0|\mu)}^2]$ to ${\cal T}^{(0)}[\mathbb{R}_{(0|\mu+\delta\mu)}^2]$,
\begin{align}\label{infinitesimalZmap2}
Z^{(\mu+\delta\mu)}(z,\zb)=Z^{(\mu)}(z,\zb)+{\delta\mu\over 2\pi^2}\int_{\mathbb{R}^2}d^2x{\bar{T}^{(\mu)}(x,\bar{x})+\mu\partial_{\mu}\bar{T}^{(\mu)}(x,\bar{x})\over z-x}+{\cal O}(\delta\mu^2)\ .
\end{align}
Note that we have two versions of the infinitesimal transformations \eqref{infinitesimalZmap} and \eqref{infinitesimalZmap2}, but we have not found a way to directly infer one from the other. However, one can check a consistency. 
The difference of the two equations \eqref{infinitesimalZmap} and \eqref{infinitesimalZmap2} yields
\begin{equation}
\begin{aligned}
\left(Z^{(0|\mu+\delta\mu)}-Z^{(\mu|\delta\mu)}\right)-\left(Z^{(0|\mu)}-Z^{(\mu|0)}\right)
&={\mu\delta\mu\over 2\pi^2}\int_{\mathbb{R}^2}d^2x{\partial_{\mu}\bar{T}^{(\mu)}(x,\bar{x})\over z-x}+{\cal O}(\delta\mu^2)\\
&\approx{\mu\over 2\pi^2}\int_{\mathbb{R}^2}d^2x{\bar{T}^{(\mu+\delta\mu)}(x,\bar{x})-\bar{T}^{(\mu)}(x,\bar{x})\over z-x}
\end{aligned}
\end{equation}
where we used $\Tb^{(\mu+\delta\mu)}=\Tb^{(\mu)}+\delta\mu\partial_{\mu}\bar{T}^{(\mu)}+{\cal O}(\delta\mu^2)$. This implies that
\begin{align}\label{master}
Z^{(0|\mu+\delta\mu)}-Z^{(\mu|\delta\mu)}={\mu\over 2\pi^2}\int d^2x{\bar{T}^{(\mu+\delta\mu)}(x,\bar{x})\over z-x}+{\cal O}(\delta\mu^2)\ .
\end{align}
We can view this relation as more fundamental than \eqref{infinitesimalZmap}, \eqref{finiteZmap}, and \eqref{infinitesimalZmap2} in the sense that  all of them can be derived from it: The $\delta\mu\to 0$ limit, in which $Z^{(\mu|0)}= Z^{(\mu)}$ and $Z^{(0|\mu)}=z$, yields \eqref{finiteZmap} which is equivalent to \eqref{infinitesimalZmap2}. Then using \eqref{finiteZmap} and \eqref{infinitesimalZmap2}, this relation yields \eqref{infinitesimalZmap}.

To avoid the clutter of notation, from now on, we drop the superscript of $Z^{(\mu)}$ and denote it simply as $Z$.
Using $\bar{\del}{1\over z-x}=2\pi\delta^2(z-x)$, the conservation $\bar{\del}T^{(\mu)}+\del\Theta^{(\mu)}=0$ and integration by parts, the finite map \eqref{finiteZmap} implies the transformation
\begin{align}\label{dZdz}
\left(
\begin{array}{c}
dZ\\
d\bar{Z}
\end{array}
\right)
=\left(
\begin{array}{cc}
1-{\mu\over\pi}\Theta^{(\mu)}(z,\zb) & {\mu\over \pi}\Tb^{(\mu)}(z,\zb) \\
{\mu\over \pi}T^{(\mu)}(z,\zb) & 1-{\mu\over\pi}\Theta^{(\mu)}(z,\zb)
\end{array}
\right)
\left(
\begin{array}{c}
dz\\
d\zb
\end{array}
\right)
\end{align}
and its inverse
\begin{align}\label{dZdzinv}
\left(
\begin{array}{c}
dz\\
d\bar{z}
\end{array}
\right)
=J^{-1}\left(
\begin{array}{cc}
1-{\mu\over\pi}\Theta^{(\mu)}(z,\zb) & -{\mu\over \pi}\Tb^{(\mu)}(z,\zb) \\
-{\mu\over \pi}T^{(\mu)}(z,\zb) & 1-{\mu\over\pi}\Theta^{(\mu)}(z,\zb)
\end{array}
\right)
\left(
\begin{array}{c}
dZ\\
d\bar{Z}
\end{array}
\right)\ ,
\end{align}
where the determinant 
\begin{align}\label{JacobianTheta}
J=1-{2\mu\over\pi}\Theta^{(\mu)}(z,\zb)-{\mu^2\over\pi^2}\left(T^{(\mu)}(z,\zb)\Tb^{(\mu)}(z,\zb)-\Theta^{(\mu)}(z,\zb)^2\right)
=1-{\mu\over\pi}\Theta^{(\mu)}(z,\zb)
\end{align}
using the flow equation \eqref{floweqn} which we will rederive from the finite map consideration in the next subsection. 
We note that the dynamical coordinate $Z$ is holomorphic, {\it i.e.}, $\partial Z/\partial\bar{Z}=0$, as can easily be checked.

\subsubsection{Mapping the stress tensors}
\label{sec:stressmap}

We now give a field-theoretic derivation of the relation between the $T\Tb$-deformed stress tensor $(T^{(\mu)}, \Tb^{(\mu)}, \Theta^{(\mu)})$ and that of the undeformed CFT on the $T\Tb$-deformed space, $(T(Z), \Tb(\bar{Z}))$. The finite version of \eqref{conservation_inf} reads
\begin{align}\label{CFTtoTTbar}
{\p\over\p x^a}T^{(\mu)}\mbox{}^a_{\mbox{ }b}(x)={\p\over\p X^a}T^{(0)}\mbox{}^a_{\mbox{ }b}(X)=0\ ,
\end{align}
where $x^a=(z,\zb)$ and $X^a=(Z,\bar{Z})$. These conservation laws yield
\begin{equation}
\begin{aligned}\label{ConservationCFTTTbar}
0&=\pb T^{(\mu)}+\p \Theta^{(\mu)}\\
&={\p\over\p \bar{Z}}T(Z)={\p z\over\p \bar{Z}}\p T(Z)+{\p\zb\over\p\bar{Z}}\pb T(Z)\\
&=J^{-1}\biggl[\pb T(Z)-{\mu\over\pi}\p\left(\Tb^{(\mu)}(z,\zb)T(Z)\right)
-{\mu\over\pi}\pb\left(\Theta^{(\mu)}(z,\zb) T(Z)\right)\biggr]\ ,
\end{aligned}
\end{equation}
where we used the conservation law $\p \Tb^{(\mu)}+\pb\Theta^{(\mu)}=0$ to obtain the last expression.
This suggests that
\begin{align}
T^{(\mu)}(z,\zb)&=T(Z)-{\mu\over\pi}\Theta^{(\mu)}(z,\zb) T(Z)\ ,\\
\Theta^{(\mu)}(z,\zb)&=-{\mu\over\pi}\Tb^{(\mu)}(z,\zb)T(Z)\ .
\end{align}
Solving these equations for $T^{(\mu)}$ and $\Theta^{(\mu)}$, we find that
\begin{align}
T^{(\mu)}(z,\zb)={T(Z)\over 1-{\mu^2\over\pi^2}\Tb(\bar{Z})T(Z)}\ ,\qquad
\Theta^{(\mu)}(z,\zb)={-{\mu\over\pi}\Tb(\bar{Z})T(Z)\over 1-{\mu^2\over\pi^2}\Tb(\bar{Z})T(Z)}\ .\label{TmuTZ}
\end{align}
These reproduce the relations derived from a holographic argument \cite{Guica:2019nzm} which we review in Appendix \ref{app:mapfromgravity}\@. The inverse maps are given by
\begin{align}\label{TTbarThetaTZ}
T(Z)={T^{(\mu)}\over 1-{\mu\over\pi}\Theta^{(\mu)}}\ ,\quad
\Tb(\bar{Z})={\Tb^{(\mu)}\over 1-{\mu\over\pi}\Theta^{(\mu)}}\ ,\quad
{\mu\over\pi}\Tb(\bar{Z})T(Z)=-{\Theta^{(\mu)}\over 1-{\mu\over\pi}\Theta^{(\mu)}}\ .
\end{align}
Note that as in the infinitesimal maps, the composite operators need to be regularized and we adopt the point-splitting regularization and renormalization. 
The consistency of these relations requires the flow equation \eqref{floweqn}
\begin{align}\label{TTbarTheta}
\Theta^{(\mu)}(z,\zb)=-{\mu\over\pi}\left[T^{(\mu)}\Tb^{(\mu)}(z,\zb)-\Theta^{(\mu)}(z,\zb)^2\right]\ .
\end{align}
Note that the maps \eqref{TmuTZ} and \eqref{TTbarThetaTZ} are of the standard form of quasi-primary operators:
\begin{align}\label{primaryProp}
{\cal O}_{\Delta,\bar{\Delta}}^{(\mu)}(x)=\left(\det{\p x\over \p X}\right)^{-{\Delta+\bar{\Delta}\over 2}}
{\cal O}_{\Delta,\bar{\Delta}}(X)
\end{align}
with $x^a=(z,\zb)$ and $X^a=(Z,\bar{Z})$ and the identifications, $T(Z)={\cal O}_{2,0}(X)$, $\Tb(\bar{Z})={\cal O}_{0,2}(X)$ and $-{\mu\over\pi}T(Z)\Tb(\bar{Z})={\cal O}_{1,1}(X)$.
As a remark, the map for the trace is equivalent to the following transformation of the $T\Tb$-operator,  $d^2x\left[T^{(\mu)}\Tb^{(\mu)}(z,\zb)-\Theta^{(\mu)}(z,\zb)^2\right]=d^2X T(Z)\Tb(\bar{Z})$.

One can, for example, explicitly check that \eqref{TTbarThetaTZ} agrees with \eqref{DeformedTformula} to second order to which the Jacobian factor in \eqref{TTbarThetaTZ} starts contributing.
Another consistency check of the infinitesimal and finite maps, \eqref{infinitesimalZmap} and  \eqref{finiteZmap}, is to see if  the recursion relation \eqref{DeformedThetaformula} and flow equation \eqref{TTbarTheta} are equivalent. The consistency requires that
\begin{equation}
\begin{aligned}\label{consistency}
\mu\p_{\mu}\left[T^{(\mu)}(z,\zb)\Tb^{(\mu)}(z,\zb)-\Theta^{(\mu)}(z,\zb)^2\right]
&=-{1\over 2\pi}\partial_z\int_{\mathbb{R}^2}d^2x{\Tb^{(\mu)}(x, \bar{x})\Theta^{(\mu)}(z,\bar{z})\over z-x}\\
&\quad -{1\over 2\pi}\partial_{\zb}\int_{\mathbb{R}^2}d^2x{T^{(\mu)}(x, \bar{x})\Theta^{(\mu)}(z,\bar{z})\over \bar{z}-\bar{x}}\ .
\end{aligned}
\end{equation}
Using \eqref{DeformedTconcise} and \eqref{DeformedThetaformula}, one can check that the LHS can indeed be rewritten as the RHS with $\Theta^{(\mu)}$ being replaced by $-{\mu\over\pi}\det T^{(\mu)}_{ij}$. We can view the consistency condition \eqref{consistency} as a flow equation for the $T\Tb$-operator:
\begin{equation}
\begin{aligned}\label{TTbarflow}
\hspace{-.2cm}
\p_{\mu}\det T^{(\mu)}_{ij}
&=\partial_z\int{d^2x\over 2\pi^2}{\Tb^{(\mu)}(x, \bar{x})\det T^{(\mu)}_{ij}(z,\zb)\over z-x}
+\partial_{\zb}\int{d^2x\over 2\pi^2}{T^{(\mu)}(x, \bar{x})\det T^{(\mu)}_{ij}(z,\zb)\over \bar{z}-\bar{x}}\ .
\end{aligned}
\end{equation}
In the following sections, we will study the implications of these maps, capitalizing on the alternative description of the $T\Tb$-deformed CFT as the undeformed CFT on the $T\Tb$-deformed space, ${\cal T}^{(0)}[\mathbb{R}_{(0|\mu)}^2]$. 

\section{The $T\Tb$-deformed space at short distances}
\label{sec:TTbardeformedspace}

In this section, we discuss the property of the $T\Tb$-deformed dynamical space that can be inferred from the finite map \eqref{finiteZmap}. 
Using \eqref{dZdzinv} and \eqref{TmuTZ}, the metric on the $T\Tb$-deformed space is given by
\begin{align}\label{Zmetric}
ds_{\mathbb{R}^2}^2=dzd\bar{z}=\left(dZ-{\mu\over \pi}\Tb(\bar{Z})d\bar{Z}\right)\left(d\bar{Z}-{\mu\over \pi}T(Z)dZ\right)
\end{align}
A remark is in order: As suggested in \cite{Guica:2019nzm} and further discussed in \cite{Caputa:2020lpa}, when the 2$d$ metric~\eqref{Zmetric} is uplifted to a 3$d$ space in a specific manner, this particular form of the metric is very suggestive of its connection to the $AdS_3$ space in the Fefferman-Graham coordinates or the Ba\~nados space \cite{Banados:1998gg}
\begin{equation}
\begin{aligned}\label{3Dbulkmetric}
ds_{3D}^2&={d\mu^2\over 4\mu^2}+{\pi \over \mu}\biggl[\left(1+{\mu^2\over\pi^2}T(Z)\Tb(\bar{Z})\right)dZd\bar{Z}
-{\mu\over\pi}T(Z)dZ^2-{\mu\over\pi}\Tb(Z)d\bar{Z}^2\biggr]\ ,
\end{aligned}
\end{equation}
where the $T\Tb$-deformation coupling $\mu$ is identified with the
radial coordinate.  This makes 
contact with the proposal in \cite{McGough:2016lol} for the holographic
dual \cite{Maldacena:1997re} of the $T\Tb$-deformed CFT as a \lq\lq cutoff'' $AdS_3$ space.  So
the undeformed CFT on the $T\Tb$-deformed space, ${\cal
T}^{(0)}[\mathbb{R}^2_{(0|\mu)}]$, is the description suited to discuss
the $AdS_3$ dual of the $T\Tb$-deformed CFT, ${\cal
T}^{(\mu)}[\mathbb{R}^2_{(\mu|0)}]$.  However, since the coordinates
$(Z, \bar{Z})$ are not ordinary coordinates but operators and depend on
the states and operator insertions, {\it i.e.}, dynamical, it is not
clear how the holography works beyond the pure gravity limit, {\it
i.e.}, in the presence of matter operators dual to bulk
fields.\footnote{\label{footHS}There is a mundane but working gravity
dual description of the $T\Tb$-deformed CFT including matter
\cite{Hirano:2020nwq}. This gravity dual can be interpreted as a
Gaussian ensemble of the AdS$_3$/CFT$_2$ in a concrete and specific
sense.} We will come back to this point later in section
\ref{sec:2ptcutoffAdS}, where we discuss that it may be possible to
include matter correlators in the \lq\lq cutoff'' $AdS_3$ if we restrict
ourselves to the semi-heavy operators ${\cal O}_{\Delta, \Delta}$, that is,
the operators of large conformal dimension $\Delta\gg \sqrt{c}$, or
alternatively, in the double scaling limit $\mu\to 0$ and
$\Delta\to\infty$ with $\mu\Delta^2$ fixed.

\begin{figure}[h!]
\centering
\includegraphics[scale=0.4]{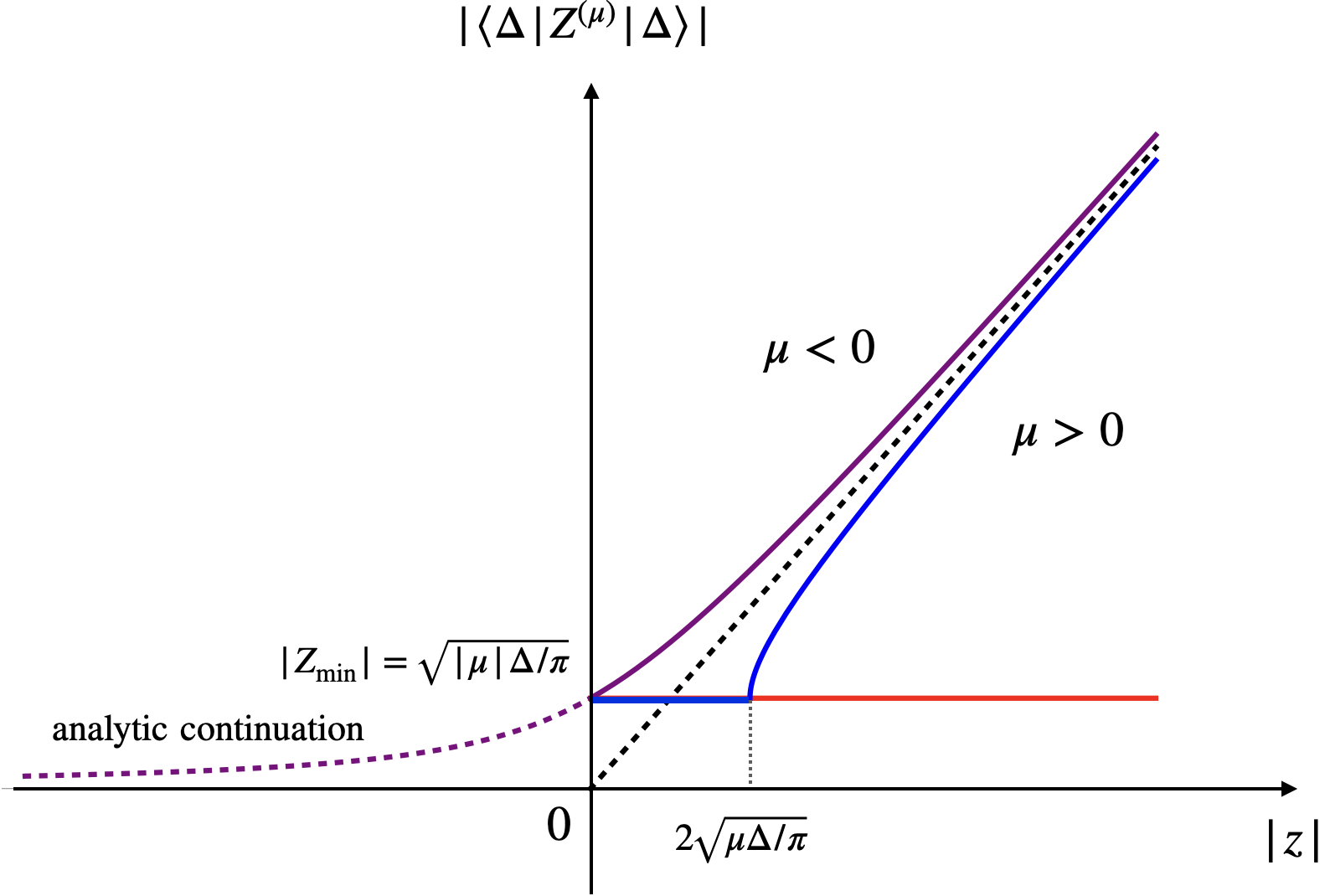}
\caption{\sl The modulus of the deformed coordinate $Z^{(\mu)}(z,\zb)\equiv Z$ on a primary state $|\Delta\rangle$. The deformed space has a state-dependent minimal length $|Z_{\rm min}|=\sqrt{\mu\Delta/\pi}$ in the UV cutoff phase $\mu>0$. The dashed black line is the undeformed map $|\langle\Delta|Z^{(\mu)}|\Delta\rangle|=|z|$ for $\mu=0$ as a reference. The solid blue curve is the map between $|z|$ and $|\langle\Delta|Z^{(\mu)}|\Delta\rangle|$ for the UV cutoff phase $\mu>0$, whereas the purple solid curve is the map for the Hagedorn phase $\mu<0$. In the latter case, the $Z$-plane can be analytically continued below the minimal length $|Z_{\rm min}|$, as indicated by the dashed purple curve, and interpreted as \lq\lq free space''. In the IR regime, $|z|\gg \sqrt{|\mu|\Delta}$, the $T\Tb$-deformation is irrelevant and the solid blue and purple curves approach the undeformed dashed black line. }
\label{fig:Zmu}
\end{figure}

\begin{figure}[h!]
\centering
\includegraphics[scale=0.4]{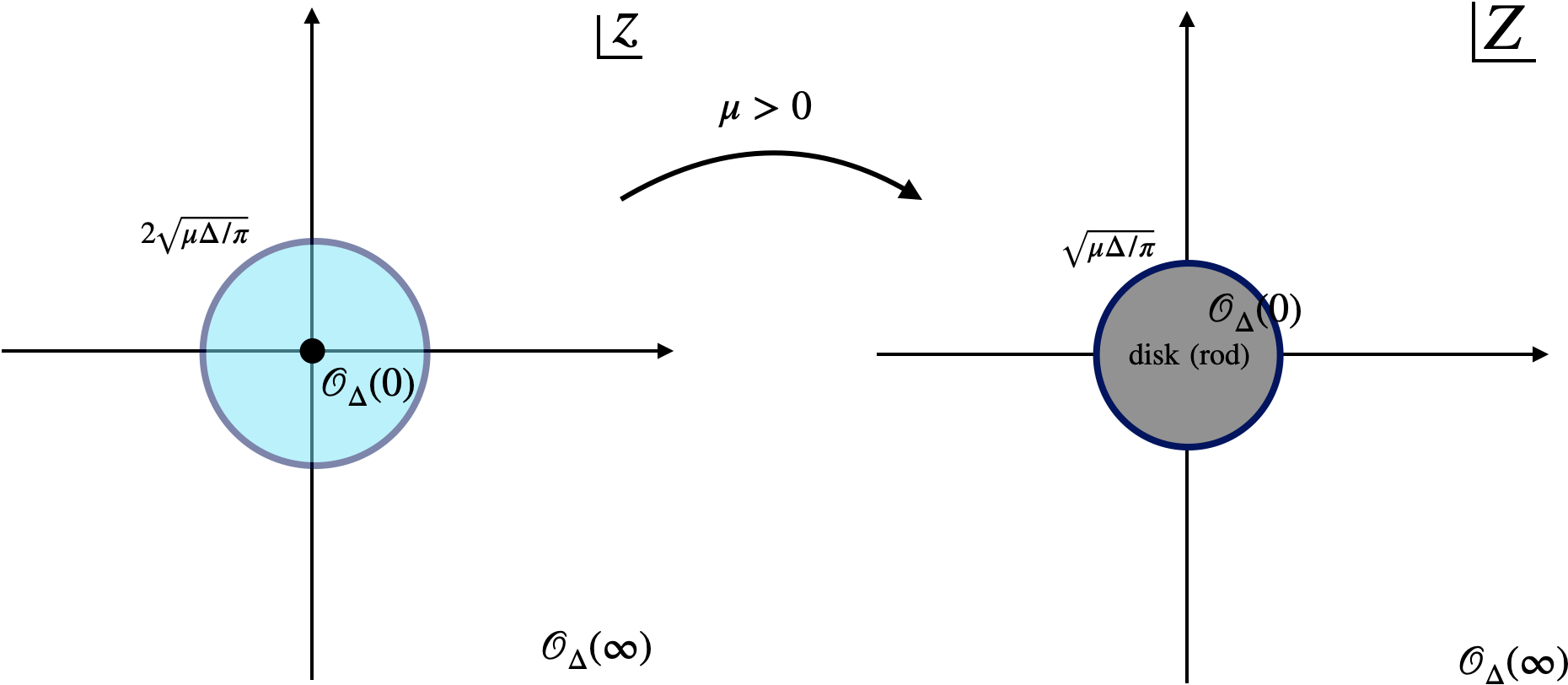}
\caption{\sl The UV cutoff phase $\mu>0$: The map between the undeformed and deformed spaces on a primary state $|\Delta\rangle$. The (light blue) disk surrounding the operator ${\cal O}_{\Delta}(0)$ in the undeformed space degenerates to a circle of radius $R=\sqrt{\mu\Delta/\pi}$ in the deformed space and the (grey) disk inside is excised. So the disk can be interpreted as a puffed-up operator analogous to a hard-rod in 1$d$ discussed in \cite{Jiang:2020nnb, Cardy:2020olv}. This is a manifestation of nonlocality in the $T\Tb$-deformed theory.}
\label{fig:plus_mu}
\end{figure}

\begin{figure}[h!]
\centering
\includegraphics[scale=0.4]{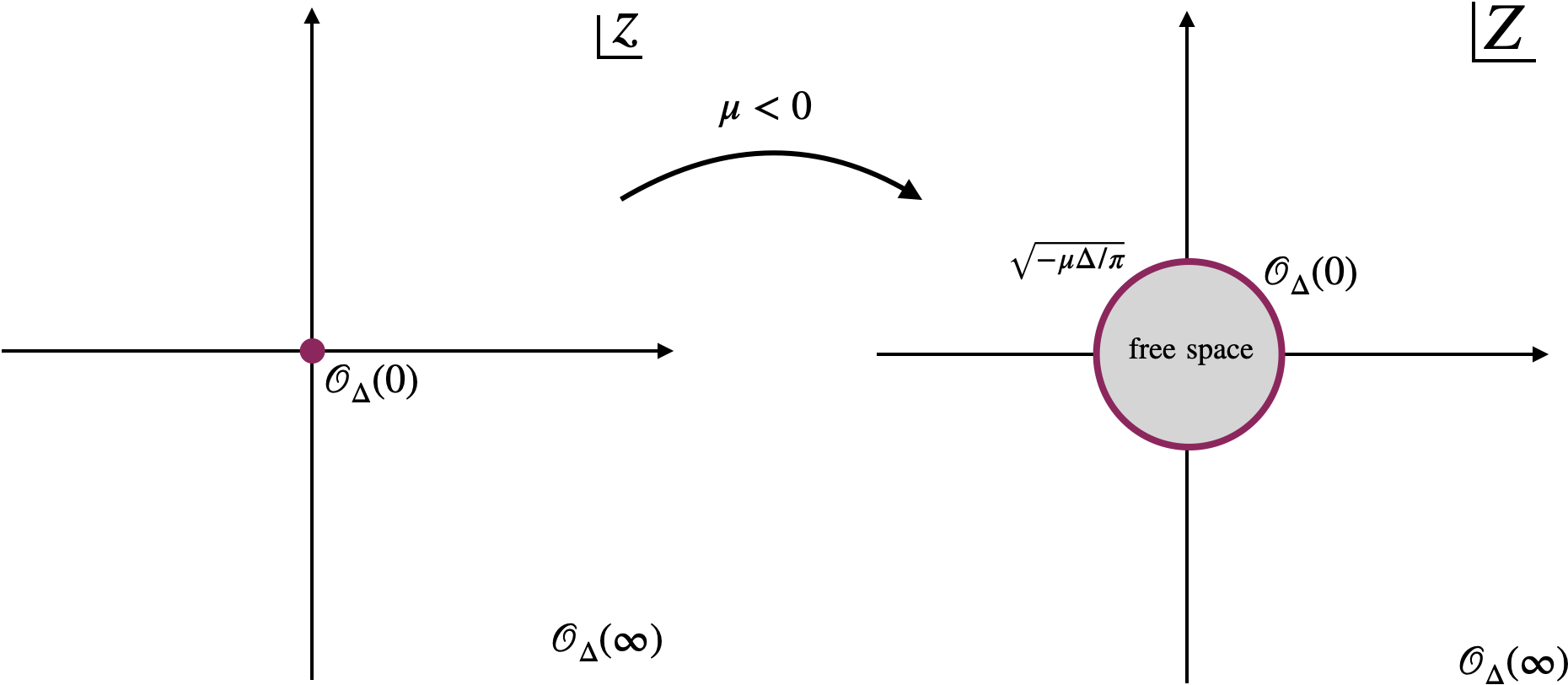}
\caption{\sl The Hagedorn phase $\mu<0$: The map between the undeformed and deformed spaces on a primary state $|\Delta\rangle$. The operator ${\cal O}_{\Delta}$ at the origin $z=0$ in the undeformed space is mapped to a circle of radius $R=\sqrt{|\mu|\Delta/\pi}$ in the deformed space. In contrast to the UV cutoff phase, no region in the $z$-plane is excised upon the map to the $Z$-plane. In fact, the $Z$-plane can be analytically continued below the minimal length $|Z_{\rm min}|$, and it looks as if free space was created ``inside'' of the operator in the analytically continued $Z$-plane as discussed in \cite{Cardy:2020olv}.}
\label{fig:minus_mu}
\end{figure}

With the coordinates $(Z, \bar{Z})$ being dynamical, there is more to be learned of the $T\Tb$-deformed space $\mathbb{R}^2_{(0|\mu)}$ than the form of the metric \eqref{Zmetric} can tell. To study how the $T\Tb$-deformed space backreacts to the state or operator insertions, we first consider the expectation value of the coordinate $Z$ on a CFT primary state $|\Delta\rangle$:
\begin{align}
\langle\Delta|Z|\Delta\rangle = z+{\mu\over 2\pi^2}\int_{\mathbb{R}^2}d^2x{\langle\Delta |\bar{T}^{(\mu)}(x,\bar{x})|\Delta\rangle\over z-x}\ .
\end{align}
The stress tensor expectation value $\langle\Delta |\bar{T}^{(\mu)}(x,\bar{x})|\Delta\rangle$ on the RHS can be computed exactly by solving the conservation law $\p\Tb^{(\mu)}+\pb\Theta^{(\mu)}=0$ and flow equation \eqref{TTbarTheta} with the ansatz
\begin{align}\label{ansatz}
\langle\Delta |\bar{T}^{(\mu)}(x,\bar{x})|\Delta\rangle={\Delta\over\bar{z}^2}f(r/\sqrt{\mu})\qquad\mbox{where}\qquad r=|z|
\end{align}
and the boundary condition $\lim_{\mu\to 0}f(r/\sqrt{\mu})=1$ in the CFT limit. One can find that
\begin{align}\label{ShigeT}
\langle\Delta|T^{(\mu)}(z,\zb)|\Delta\rangle ={\Delta\over z^2\sqrt{1-{4\mu\Delta\over \pi|z|^2}}}\ ,\qquad
\langle\Delta|\Theta^{(\mu)}(z,\zb)|\Delta\rangle=-{\pi\over 4\mu}{\left(1-\sqrt{1-{4\mu\Delta\over\pi |z|^2}}\right)^2\over \sqrt{1-{4\mu\Delta\over\pi |z|^2}}}\ .
\end{align}
We checked this result against the explicit form of the stress tensor deformation to second order given in Appendix \ref{app:secondorder}\@.
Note that to express $\langle\Delta|\Theta^{(\mu)}|\Delta\rangle$ in terms of the function $f(r/\sqrt{\mu})$, the factorization property of the $T\Tb$-operator \cite{Zamolodchikov:2004ce} is used:
\be
\langle\Delta|T^{(\mu)}\Tb^{(\mu)}-(\Theta^{(\mu)})^2|\Delta\rangle
=\langle\Delta|T^{(\mu)}|\Delta\rangle \langle\Delta|\Tb^{(\mu)}|\Delta\rangle-(\langle\Delta|\Theta^{(\mu)}|\Delta\rangle)^2\ .
\ee
The stress tensor expectation value then yields\footnote{By integration by parts, the integral can be performed as
\begin{equation}
\begin{aligned}
\int_{\mathbb{R}^2} {d^2x\over (z-x)\bar{x}^2\sqrt{1-{4\mu\Delta\over\pi|x|^2}}}
&={\pi\over 2\mu\Delta}\int_{\mathbb{R}^2} {d^2x x\over (z-x)}\bar{\partial}\sqrt{1-{4\mu\Delta\over\pi|x|^2}}
={\pi^2z\over \mu\Delta}\left[\sqrt{1-{4\mu\Delta\over\pi |z|^2}}-1\right]\ ,
\end{aligned}
\end{equation}
where we used $\bar{\partial}_x(1/(z-x))=-2\pi\delta^2(z-x)$.}
\begin{equation}
\begin{aligned}\label{Zmuvev}
\langle \Delta|Z|\Delta\rangle={z\over 2}\left[\sqrt{1-{4\mu\Delta\over\pi |z|^2}}+1\right]\ .
\end{aligned}
\end{equation}
The modulus $|\langle \Delta|Z|\Delta\rangle|=|\langle \Delta|Z^{(\mu)}|\Delta\rangle|$ is plotted in Figure \ref{fig:Zmu}. The property of the $T\Tb$-deformed space differs qualitatively depending on the sign of the $T\Tb$ coupling $\mu$.
In the case of the UV cutoff phase $\mu>0$,\footnote{The usage of the terminology, UV cutoff phase for $\mu>0$ and Hagedorn phase for $\mu<0$, comes from the behavior of the energy spectrum and free energy of the $T\Tb$-deformed CFT on a cylinder \cite{Zamolodchikov:2004ce,Smirnov:2016lqw, Cavaglia:2016oda, Cardy:2018sdv}: $E={\pi R\over \mu}(1-\sqrt{1-2\mu(\Delta+\bar{\Delta}-c/12)/(\pi R^2)}$ and $F(\beta)={1\over 2\mu}(1-\sqrt{1+2\pi c\mu/(3\beta^2)})$, where $R$ is the radius of (spatial) circle and $\beta$ is the inverse temperature. The former implies that the $T\Tb$-deformed energy for $\mu>0$ becomes complex at high energies suggesting presumably a UV cutoff, whereas the latter implies that there is a limiting (Hagedorn) inverse temperature $\beta_H=\sqrt{-2\pi c\mu/3}$ for $\mu<0$ below which the free energy becomes complex.} there appears a state-dependent minimal length in the deformed space $\mathbb{R}_{(0|\mu)}$:
\begin{align}\label{minradius}
|\langle \Delta|Z|\Delta\rangle|\ge \sqrt{\mu\Delta\over\pi} \equiv |Z_{\rm min}|\ ,
\end{align}
since the modulus stays constant below $|z|\le 2\sqrt{\mu\Delta/\pi}$. So the backreaction of an operator of dimension $\Delta$ cuts off a disk of radius $\sqrt{\mu\Delta/\pi}$, or put differently, the operator puffs up into a disk in the deformed space and becomes a non-local object.
This is a manifestation of the hard-rod picture discussed in \cite{Jiang:2020nnb, Cardy:2020olv} as illustrated in Figure \ref{fig:plus_mu}.
In contrast, in the Hagedorn phase $\mu<0$, there is no minimal length. Rather, it is as if the space is enlarged in comparison to the undeformed $\mathbb{R}^2$ as indicated by the dashed purple curve in Figure \ref{fig:Zmu}. The map \eqref{Zmuvev} suggests that, from the perspective of the deformed space, the undeformed $\mathbb{R}^2$ can be smoothly extended into the region $r=|z|<0$:
\begin{align}\label{Zminuscontinue}
|\langle \Delta|Z|\Delta\rangle|=\left\{
\begin{array}{cl}
{r\over 2}\left[1+\sqrt{1+{4|\mu|\Delta\over\pi r^2}}\right] & \quad\mbox{for}\quad r \ge 0\\
{r\over 2}\left[1-\sqrt{1+{4|\mu|\Delta\over\pi r^2}}\right] & \quad\mbox{for}\quad r < 0\quad ({\rm analytic\,\,continuation})
\end{array}
\right.\ .
\end{align} 
This enlarged region is the free space discussed in \cite{Cardy:2020olv}. So the backreaction of an operator of dimension $\Delta$ creates a free space inside the disk of radius $\sqrt{|\mu|\Delta/\pi}$, or put differently, the operator becomes a ring that houses free space inside of it as illustrated in Figure \ref{fig:minus_mu}.

A few remarks are in order:  (1) Curiously, the expectation value of the $T\Tb$-deformed coordinate satisfies the following inviscid Burgers' equation:
\begin{align}
\p_{\mu}\left(\mu^{-1}|\langle\Delta|Z|\Delta\rangle|\right)
=-\left(\mu^{-1}|\langle\Delta|Z|\Delta\rangle|\right){\p\over\p |z|}\left(\mu^{-1}|\langle\Delta|Z|\Delta\rangle|\right)
\end{align}
in a similar way to the energy of the $T\Tb$-deformed theory on a cylinder \cite{Smirnov:2016lqw}.
(2) We observe that 
\begin{align}\label{TZvev}
\langle\Delta|T(Z)|\Delta\rangle={\langle\Delta|T^{(\mu)}|\Delta\rangle\over 1-{\mu\over\pi}\langle\Delta|\Theta^{(\mu)}|\Delta\rangle}
={\Delta\over (\langle\Delta|Z|\Delta\rangle)^2}
\end{align}
where we assumed the factorization property $\langle\Delta|T^{(\mu)}(\Theta^{(\mu)})^n|\Delta\rangle=\langle\Delta|T^{(\mu)}|\Delta\rangle(\langle\Delta|\Theta^{(\mu)}|\Delta\rangle)^n$. This is of the form of the CFT stress tensor expectation value as one might have expected.
The expression of the dynamical coordinate expectation value \eqref{Zmuvev} must be consistent with the coordinate map \eqref{Zmetric} between $(z, \zb)$ and $(Z, \bar{Z})$. Indeed, with \eqref{TZvev}, we see that the inverse map of \eqref{Zmuvev} is nothing but the coordinate map \eqref{Zmetric}:
\begin{align}
z=\langle\Delta |Z|\Delta\rangle+{\mu\Delta\over \pi \langle\Delta|\bar{Z}|\Delta\rangle}
=\int\left(d\langle\Delta |Z|\Delta\rangle-{\mu\over\pi}\langle\Delta|\Tb(\bar{Z})|\Delta\rangle d\langle\Delta|\bar{Z}|\Delta\rangle\right)\ .
\end{align}
In this form, the analytic continuation \eqref{Zminuscontinue} in the Hagedorn phase $\mu<0$ is naturally covered.
(3) The circle of radius $\sqrt{|\mu|\Delta/\pi}$ corresponds to the coordinate singularity of the metric \eqref{Zmetric}:
\begin{align}
\det g_{ab}(Z,\bar{Z})=-{1\over 4}\left(1-{\mu^2\over\pi^2}T(Z)\Tb(\bar{Z})\right)^2=0\ .
\end{align}
On the state $|\Delta\rangle$, this yields $|\langle\Delta|Z|\Delta\rangle|=\sqrt{|\mu|\Delta/\pi}$. 

\begin{figure}[h!]
\centering
\includegraphics[scale=0.25]{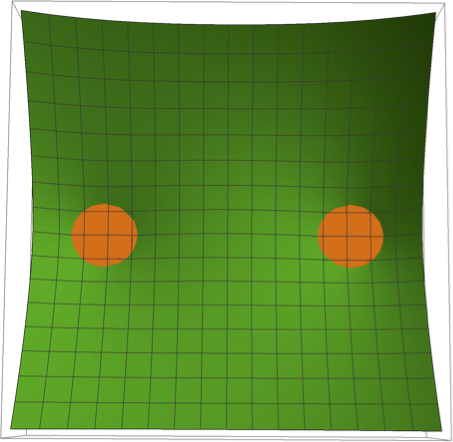}\hspace{3cm}
\includegraphics[scale=0.25]{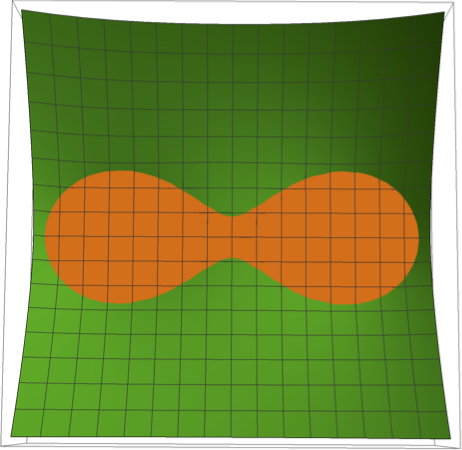}
\caption{\sl The backreaction of two operator insertions in the $Z$-plane: The orange regions indicate the cut-off region ($\mu>0$) or free space $(\mu<0)$. On the left, the two operators of dimension $\Delta$ are well separated at a distance $2a$ and each orange region is approximately a disk of radius $\sqrt{|\mu|\Delta/\pi}\ll a$. On the right, as the two operators come closer to each other, $a\sim \sqrt{|\mu|\Delta/\pi}$, the two disks merge and the individual (fat) operator cannot be resolved.}
\label{fig:2pttop}
\end{figure}

This last observation gives us a simple way to generalize the analysis for a single operator insertion to multiple operator insertions. In the case of two operator insertions at $Z=-a$ and $Z=a$,  the CFT stress tensor is given by
\begin{align}
T(Z)={(2a)^2\Delta\over (Z-a)^2(Z+a)^2}\ ,
\label{T(Z)2pt}
\end{align}
where $Z$ and $T(Z)$ should be understood as expectation values. So the boundary of the cut-off region or free space reads
\begin{align}
|(Z/a-1)(Z/a+1)|={2\over a}\sqrt{|\mu|\Delta/\pi}\ .
\end{align}
This is illustrated in Figure \ref{fig:2pttop}. As the two fat operators approach each other to the distance scale shorter than their sizes, they merge and the individual operator cannot be resolved. This is yet another manifestation of non-locality of the $T\Tb$-deformed theory.

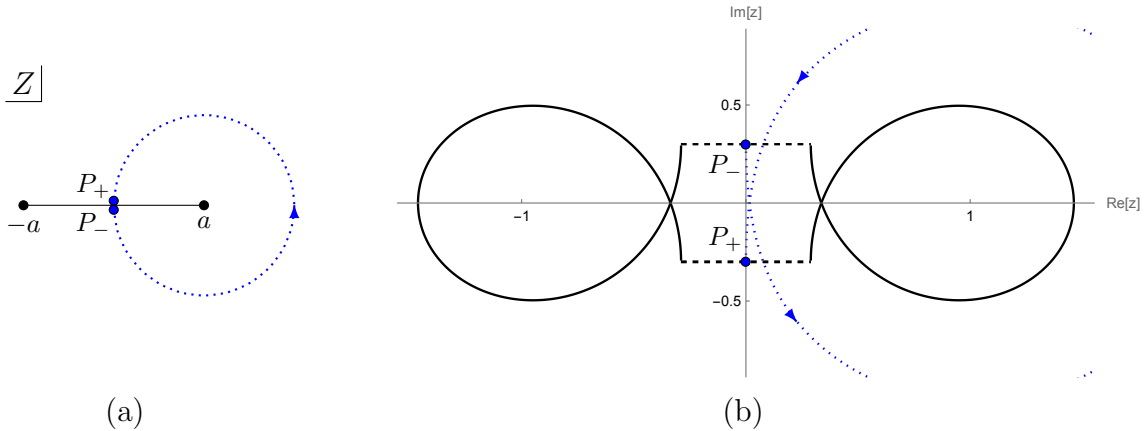
\begin{figure}[h]
 \begin{center}
\begin{tabular}{ccc}
  \raisebox{1.1cm}{%
 \begin{tikzpicture}[scale=1.2]
  \draw (-1.2,1.15) -- +(0.4,0) -- +(0.4,0.4);
 \node () at (-1.0,1.35) {$Z$};
 \draw (-1,0) -- (1,0);
 \draw [smooth,thick,dotted,blue,domain=-178:178]  plot ({1+1*cos(\x)}, {1*sin(\x)});
  \draw [thick,blue,-latex] (2,0) -- +(0,0.01);
 \draw[fill=black] (-1,0) circle (0.05) node [below] {\footnotesize $-a$};
 \draw[fill=black] ( 1,0) circle (0.05) node [below] {\footnotesize $ a$};
 \draw[fill=blue] (0, 0.05) circle (0.05) node [above left,xshift=3,yshift=-3] {\footnotesize $P_+$};
 \draw[fill=blue] (0,-0.05) circle (0.05) node [below left,xshift=3,yshift= 3] {\footnotesize $P_-$};
 \end{tikzpicture}}
 &\hspace*{0.5cm}&
 \begin{tikzpicture}[scale=1.2]
  \node[inner sep=0pt] (plot) at (0,0)
    {\includegraphics[height=5cm]{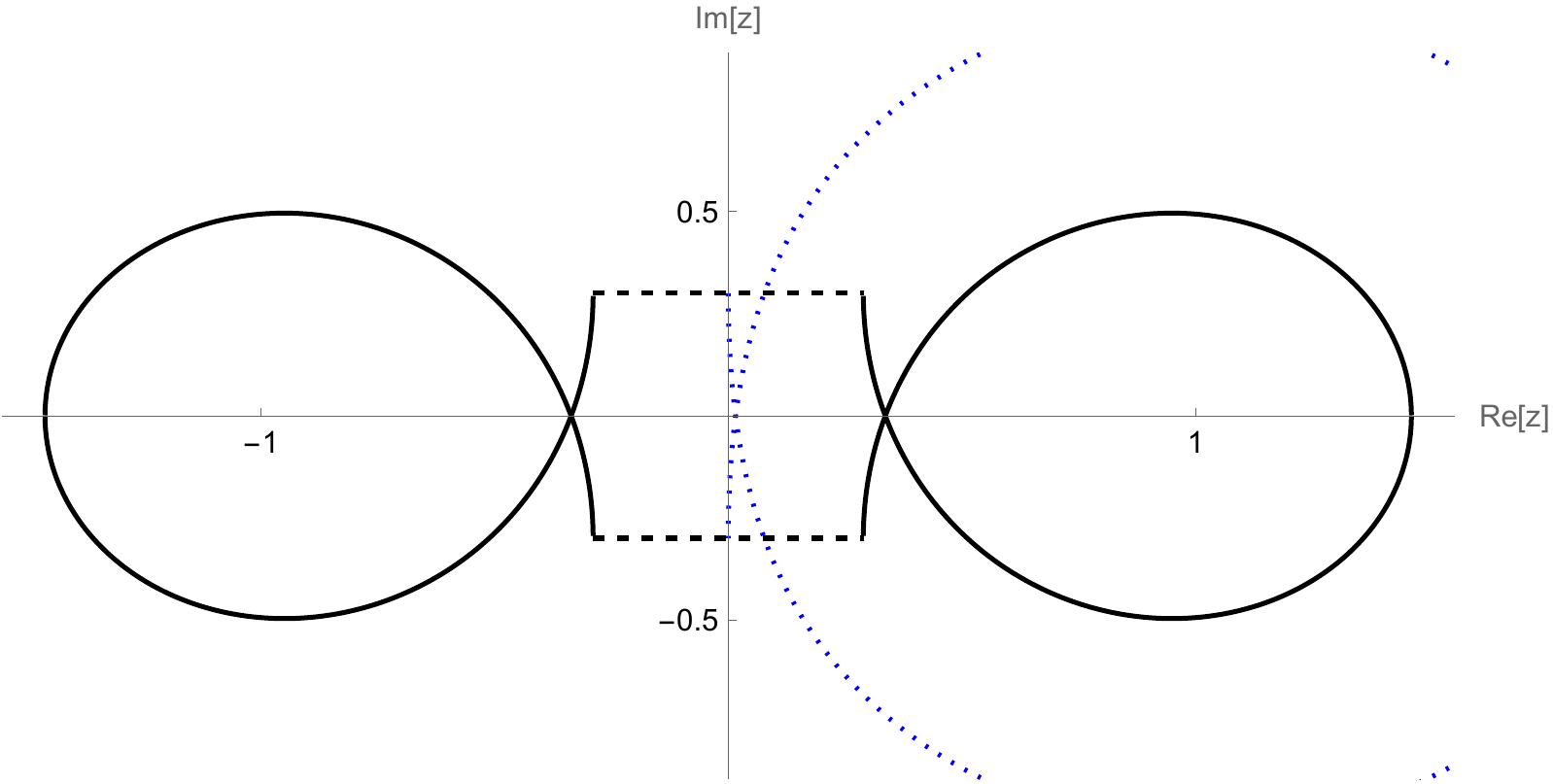}};
  \draw[fill=blue] (-0.26,0.52) circle (0.05) node [below left,xshift=3] {\footnotesize $P_-$};
  \draw[fill=blue] (-0.26,-0.78) circle (0.05) node [above left,xshift=3] {\footnotesize $P_+$};
  \draw [thick,blue,-latex] (0.31,1.2) -- +(-130:0.01);
  \draw [thick,blue,latex-] (0.31,-1.45) -- +(+130:0.01);
\end{tikzpicture}
 \\
 (a)~~~~~& &(b)~~~~~
\end{tabular}  
\caption{\sl (a) A circular path (blue dotted circle) to go around $Z=a$
   in the $Z$-space .  $P_-$ is a point below the cut $[-a,a]$ and $P_+$
   is a point above the cut, but they are really the same point. (b) The
   $z$-space.  The regions inside the black solid curves are excluded.
   The two horizontal black dashed lines are identified. A part of the
   $z$-space image of the circular path is also shown as a blue dotted
   curve. Points $P_-$ and $P_+$ are to be identified. The parameters
   are $a=\Delta=1,\mu=0.3>0$.  \label{fig:going_around_Z=a}}
 \end{center}
\end{figure}

The above demonstrates that, because of the non-trivial mapping between
the $z$-space and the dynamical $Z$-space, some regions near operator
insertions must be removed, in both spaces.  Actually, in the presence
of two insertions which makes the space non-simply connected, there is
another interesting phenomenon.  Eq.~\eqref{Zmetric} implies that the
relation between $z$ and $Z$ is
\begin{align}
 dz=dZ-{\mu\over \pi}\Tb(\Zb)d\Zb,
\end{align}
which for two insertions \eqref{T(Z)2pt} gives
\begin{align}
\label{z_vs_Z_2pt}
 z=Z+{\mu\Delta \over \pi}\left(\frac{1}{\Zb-a}+\frac{1}{\Zb+a}+\frac{1}{a}\log\frac{\Zb-a}{\Zb+a}\right).
\end{align}
The constant of integration was chosen so that $z=Z$ at infinity.  This
function has a cut between $Z=\pm a$ on the $Z$-plane.  Let us consider
moving on the $Z$-plane counterclockwise around $Z=a$ from point $P_-$
just below the cut to point $P_+$ just above the cut, as in
Figure~\ref{fig:going_around_Z=a}(a). From~\eqref{z_vs_Z_2pt}, we see that
the deformed-theory coordinate $z$ changes by
\begin{align}
 z_{P_+}^{}-z_{P_-}^{}=-{2i \mu \Delta\over  a}.
\end{align}
Namely, the two points $P_-,P_+$, which really are the same point, are
mapped in the $z$-space to two points with different values of $\Im z$.
To identify $z_{P_+}$ and $z_{P_-}$ in the $z$-space, we have to
identify the region above the cut and the region below the cut with a
vertical shift.  For $\mu<0$ the shift is $\Im z_{P_+}^{}-\Im
z_{P_-}^{}=-2\mu \Delta/a>0$ and there is a ``gap'' between the cuts,
while for $\mu>0$ the shift is $-2\mu \Delta/a<0$ and there is an
``anti-gap'' between the cuts; see Figure~\ref{fig:going_around_Z=a}(b)
for a graphical explanation for the $\mu>0$ case (with an
anti-gap). Therefore, the $z$-space is not just a copy of flat
$\mathbb{R}^2$ with two disks removed, although its metric is locally
flat.\footnote{The fact that the \TTbar\ deformation inserts shifts of
coordinates between two points has already been noted in
\cite{Cardy:2019qao} in perturbation theory.  This is a manifestation of
that phenomenon in a finite setting.}

\section{Correlators on the $T\Tb$-deformed space}
\label{sec:correlators}

The description of the $T\Tb$-deformed CFT as the CFT on the $T\Tb$-deformed space,  ${\cal T}^{(0)}[\mathbb{R}_{(0|\mu)}^2]$, may provide, both practically and conceptually, useful perspectives in the study of correlators in the $T\Tb$-deformed CFT\@.\footnote{The correlators in the $T\Tb$-deformed space  were recently studied in \cite{Aharony:2023dod} using the path integral formulation in the topological or massive gravity description \cite{Dubovsky:2012wk, Dubovsky:2017cnj,Dubovsky:2018bmo,Tolley:2019nmm}. Our study may provide a complementary view on this subject.}
 To begin with, as discussed in \cite{Cardy:2019qao},  in considering the correlators, we can take two viewpoints, ``Heisenberg'' and ``Schr\"odinger'' in analogy to quantum mechanics. In the ``Heisenberg'' picture, the operators are deformed and denoted by ${\cal O}^{(\mu)}_{\Delta, \bar{\Delta}}(z, \zb)$, whereas in the Schr\"odinger'' picture, the states are deformed:
\begin{align}\label{deformedcorrelators}
\langle {\cal O}^{(\mu)}_{\Delta_1, \bar{\Delta}_1}(z_1, \zb_1)\cdots {\cal O}^{(\mu)}_{\Delta_n, \bar{\Delta}_n}(z_n, \zb_n)\rangle_{0}=
\langle {\cal O}^{(0)}_{\Delta_1, \bar{\Delta}_1}(z_1, \zb_1)\cdots {\cal O}^{(0)}_{\Delta_n, \bar{\Delta}_n}(z_n, \zb_n)\rangle_{\mu}\ ,
\end{align}
where the subscript of $\langle\cdots\rangle_{\mu}$ denotes that the correlators are evaluated on the vacuum of the $T\Tb$-deformed theory with the coupling $\mu$. So
$\mu=0$ corresponds to the undeformed CFT vacuum. It is the ``Heisenberg'' picture on the LHS that is better suited to our purposes. As alluded to in \eqref{primaryProp}, via the coordinate transformation $(z,\zb)\mapsto (Z, \bar{Z})$, the deformed operators may be given by
\begin{align}\label{primaryoperator_map}
{\cal O}_{\Delta,\bar{\Delta}}^{(\mu)}(z,\zb)=\left(\det{\p x\over \p X}\right)^{-{\Delta+\bar{\Delta}\over 2}}
{\cal O}_{\Delta,\bar{\Delta}}(X)
=\left(1-{\mu^2\over\pi^2}T(Z)\Tb(\bar{Z})\right)^{-{\Delta+\bar{\Delta}\over 2}}
{\cal O}_{\Delta}(Z){\cal O}_{\bar{\Delta}}(\bar{Z})
\end{align}
with $x^a=(z,\zb)$ and $X^a=(Z,\bar{Z})$, which generalizes the transformation property of the stress tensor discussed in Section \ref{sec:stressmap}. We note that in the spirit of Section \ref{sec:stressdef}, the infinitesimal version of this operator map reads
\begin{align} 
{\cal O}_{\Delta,\bar{\Delta}}^{(\mu+\delta\mu)}(x)=\left(\det{\p x\over \p \tilde{x}}\right)^{-{\Delta+\bar{\Delta}\over 2}}{\cal O}_{\Delta,\bar{\Delta}}^{(\mu)}(\tilde{x})\ ,
\end{align}
where $x^a=(z,\zb)$ and $\tilde{x}^a=(Z^{(\mu|\delta\mu)}, \bar{Z}^{(\mu|\delta\mu)})$ with the latter defined in \eqref{infinitesimalZmap}. This can be expressed as the flow equation:
\begin{equation}
\begin{aligned}\label{primaryoperator_infmap}
\p_{\mu}{\cal O}_{\Delta,\bar{\Delta}}^{(\mu)}(x)&={1\over 2\pi^2}\int d^2y{\Tb^{(\mu)}(y,\bar{y})\p{\cal O}_{\Delta,\bar{\Delta}}^{(\mu)}(x)\over z-y}
+{1\over 2\pi^2}\int d^2y{T^{(\mu)}(y,\bar{y})\pb {\cal O}_{\Delta,\bar{\Delta}}^{(\mu)}(x)\over \zb-\bar{y}}\\
&-{\Delta+\bar{\Delta}\over\pi}\Theta^{(\mu)}(x){\cal O}_{\Delta,\bar{\Delta}}^{(\mu)}(x)\ .
\end{aligned}
\end{equation}
The second line is the contribution from the Jacobian which is absent in the formula proposed in \cite{Cardy:2019qao}. We believe that the appearance of a similar contribution in the stress tensor component \eqref{Tinf_flow} provides evidence for the presence of this term. 
As in the case of the stress tensor, both the finite and infinitesimal maps, \eqref{primaryoperator_map} and \eqref{primaryoperator_infmap}, require regularization and we adopt the point-splitting regularization and renormalization. 

As a demonstration of utility of \eqref{deformedcorrelators} and \eqref{primaryoperator_map}, we calculate the two-point correlator to first order in the $T\Tb$-coupling $\mu$. Since the Jacobian factor only contributes from the second order, we have
\begin{align}\label{2pt}
\langle  {\cal O}^{(\mu)}_{\Delta, \Delta}(z_1, \zb_1) {\cal O}^{(\mu)}_{\Delta, \Delta}(z_2, \zb_2)\rangle_0
=\langle {\cal O}_{\Delta}(Z_1){\cal O}_{\Delta}(\bar{Z}_1){\cal O}_{\Delta}(Z_2){\cal O}_{\Delta}(\bar{Z}_2)\rangle_0+{\cal O}(\mu^2)\ .
\end{align}
An important note is that despite the fact that ${\cal O}_{\Delta}(Z)$ and ${\cal O}_{\Delta}(\bar{Z})$ are CFT operators in disguise, the holomorphic and anti-holomorphic parts do not factorize:
\begin{align}\label{nodecoupling}
\langle {\cal O}_{\Delta}(Z_1){\cal O}_{\Delta}(\bar{Z}_1){\cal O}_{\Delta}(Z_2){\cal O}_{\Delta}(\bar{Z}_2)\rangle_0\ne 
\langle {\cal O}_{\Delta}(Z_1){\cal O}_{\Delta}(Z_2)\rangle_0\langle{\cal O}_{\Delta}(\bar{Z}_1){\cal O}_{\Delta}(\bar{Z}_2)\rangle_0\ .
\end{align}
In fact, the first-order correction solely comes from the coupling between them. Expanding ${\cal O}_{\Delta}(Z)$ in powers of $\mu$ using \eqref{finiteZmap} and \eqref{T2ndorder}, it reads
\begin{equation}
\begin{aligned}\label{2pt1storder}
{\rm RHS\,of\,}\eqref{2pt}&={1\over |z_{12}|^{4\Delta}}
-{\mu\Delta\over\pi^2z_{12}^{2\Delta+1}}\int d^2x{\langle\Tb(\bar{x}){\cal O}_{\Delta}(\zb_1){\cal O}_{\Delta}(\zb_2)\rangle_0\over  z_1-x}
+(z_1\leftrightarrow z_2)+{\rm c.c.}\\
&={1\over |z_{12}|^{4\Delta}}-{8\mu\Delta^2\over\pi |z_{12}|^{4\Delta}} \left({\ln|z_{12}/\epsilon|^2\over |z_{12}|^2}-{1\over |z_{12}|^2}-{1\over 2\epsilon}\left({1\over z_{12}}+{1\over \zb_{12}}\right)\right)
\end{aligned}
\end{equation}
to first order in $\mu$ where we used $\langle\Tb(x){\cal O}_{\Delta}(\zb_1){\cal O}_{\Delta}(\zb_2)\rangle_0=\Delta/(\zb_{12}^{2(\Delta-1)}(\bar{x}-\zb_1)^2(\bar{x}-\zb_2)^2)$ and performed a point-splitting regularization.\footnote{To provide a little more detail, we used that
\begin{align}
{1\over (\bar{x}-\zb_1)^2 (\bar{x}-\zb_2)^2}=-{1\over\zb_{12}^2}\left[\pb_x{1\over \bar{x}-\zb_1}+{2\over\zb_{12}}\pb_x\ln{\bar{x}-\zb_1\over \bar{x}-\zb_2}+\pb_x{1\over \bar{x}-\zb_2}\right]
\end{align}
and integration by parts while regularizing $1/(z_1-x)\to 1/(z_1+\epsilon-x)$.} This reproduces the results in \cite{Kraus:2018xrn, Cardy:2019qao}. Note that the computational detail is rather different from that of the conformal perturbation theory in \cite{Kraus:2018xrn} which involves an insertion of the $T\Tb$-operator, $\mu\int d^2x T(x)\Tb(\bar{x})$, in contrast to a single $T$ or $\Tb$ in this method.\footnote{Recall the definition of the $T\Tb$-deformation \eqref{infinitesimal_deformed_action}. The naive conformal perturbation theory only works to first order except for a few accidental cases since the $T\Tb$-operator itself receives higher-order corrections, whereas this method works to an arbitrary order.} Though tedious, we can systematically compute higher-order corrections by keeping on expanding ${\cal O}_{\Delta}(Z)$ and using, for example, \eqref{DeformedTformula} to find the higher-order deformation of $T^{(\mu)}$. Indeed, we have tested \eqref{primaryoperator_map} to second order, reproducing the leading-log contribution to the two-point correlator \cite{Cardy:2019qao}:
\begin{equation}
\begin{aligned}\label{2pt2ndorder}
\left\langle{\cal O}^{(\mu)}_{\Delta, \Delta}(z_1,\zb_1){\cal O}^{(\mu)}_{\Delta, \Delta}(z_2, \zb_2)\right\rangle_0
&=\cdots +{8\mu^2\over \pi^2} \Delta^2(2\Delta+1)^2{\ln^2|z_{12}/\epsilon|^2\over |z_{12}|^{4(\Delta+1)}}+\cdots\ .
\end{aligned} 
\end{equation}
We omit the computational detail as it is straightforward and not particularly illuminating. 

Even though the correlators of the $T\Tb$-deformed theory necessarily involve both holomorphic and anti-holomorphic operators $(T(Z), {\cal O}_{\Delta}(Z))$ and $(\Tb(\bar{Z}), {\cal O}_{\Delta}(\bar{Z}))$ and they do not factorize, one may still wonder if the standard CFT properties apply to (anti-)holomorphic correlators. For example, let us consider the two-point correlator and ask if the following is true:
\begin{align}\label{holomorphicEx}
\langle {\cal O}_{\Delta}(Z_1){\cal O}_{\Delta}(Z_2)\rangle_0=\left\langle {1\over (Z_1-Z_2)^{2\Delta}}\right\rangle_0\ .
\end{align}
It is not our purpose in this discussion to prove it, but we only want to show how it may be holding true in a somewhat nontrivial way. To illustrate it, we check it to second order in $\mu$.
Recall that in section \ref{sec:TTbardeformedspace}, we found $\langle Z\rangle_0=z$, as inferred from \eqref{Zmuvev}. One might then think that the dynamical coordinates $Z_i$'s are simply replaced by the flat coordinates $z_i$'s on the RHS of this equation. If so, \eqref{holomorphicEx} would not be true. But that is not the case as we now demonstrate. First, by expanding ${\cal O}_{\Delta}(Z_i)$, the LHS of the first equation can be calculated as
\begin{equation}
\begin{aligned}
\langle{\cal O}_{\Delta}(Z_1){\cal O}_{\Delta}(Z_2)\rangle_0
&={1\over z_{12}^{2\Delta}}
+{\mu^2\over 4\pi^4}\int{d^2xd^2y\langle \Tb(\bar{x})\Tb(\bar{y})\partial{\cal O}_{\Delta}(z_1)\partial{\cal O}_{\Delta}(z_2)\rangle_0\over (z_1-x)(z_2-y)}+\cdots\\
&={1\over z_{12}^{2\Delta}}-{\Delta(2\Delta+1)c\mu^2\over 4\pi^4z_{12}^{2\Delta+2}}\int {d^2xd^2y\over (z_1-x)(z_2-y)(\bar{x}-\bar{y})^4}+\cdots\\
&={1\over z_{12}^{2\Delta}}+{\Delta(2\Delta+1)c\mu^2\over 6\pi^2z_{12}^{2\Delta+2}\bar{z}_{12}^2}+{\cal O}(\mu^3)\ .
\end{aligned}
\end{equation}
where we used that $\langle T(z)\rangle_0=\langle\Tb(z)\rangle_0=0$ and integration by parts with $(\bar{x}-\bar{y})^{-4}=-\pb_x\pb_y(\bar{x}-\bar{y})^{-2}/6$. 
Meanwhile, by expanding $Z_i$'s, the RHS reads
\begin{equation}
\begin{aligned}\label{2ptZ2nd}
\left\langle{1\over Z_{12}^{2\Delta}}\right\rangle_0
&={1\over z_{12}^{2\Delta}}+{\Delta(2\Delta+1)\mu^2\over 4\pi^4z_{12}^{2\Delta+2}}\left
\langle\left(\int d^2x {\Tb(\bar{x})\over z_1-x}-\int d^2x {\Tb(\bar{x})\over z_2-x}\right)^2 \right\rangle_0+\cdots\\
&={1\over z_{12}^{2\Delta}}-{\Delta(2\Delta+1)\mu^2\over 2\pi^4z_{12}^{2\Delta+2}}\int {d^2xd^2y\langle \Tb(\bar{x})\Tb(\bar{y})\rangle_0\over (z_1-x)(z_2-y)}+{\rm div}+\cdots\\
&={1\over z_{12}^{2\Delta}}+{\Delta(2\Delta+1)c\mu^2\over 6\pi^2z_{12}^{2\Delta+2}\bar{z}_{12}^2}+{\rm div}+{\cal O}(\mu^3)\ .
\end{aligned}
\end{equation}
where we once again used that $\langle T(z)\rangle_0=\langle\Tb(z)\rangle_0=0$ and div denotes the $1/\epsilon^2$ divergent contribution from
$\int d^2xd^2y\langle \Tb(\bar{x})\Tb(\bar{y})\rangle_0/((z_i-x)(z_i-y))$ $(i=1,2)$ which we regularize as $(z_i-x)(z_i-y)\to (z_i+\epsilon-x)(z_i-y)$  by point-splitting and is renormalized away.
So we see that the two indeed agree, at least to second order, after renormalization.

As another example, let us check if the following is true to second order in $\mu$:
\begin{align}\label{TTholomorphic}
\langle T(Z_1)T(Z_2)\rangle_0=\left\langle {c\over 2(Z_1-Z_2)^4}\right\rangle_0\ .
\end{align}
By expanding $T(Z_i)$, the LHS can be calculated as
\begin{equation}
\begin{aligned}
\left\langle T(Z_1)T(Z_2)\right\rangle_0 
&={c\over 2z_{12}^4}
+{\mu^2\over 4\pi^4}\int d^2xd^2y {\langle\Tb(\bar{x})\Tb(\bar{y})\rangle_0\langle \p T(z_1)\p T(z_2)\rangle_0\over (z_1-x)(z_2-y)}\\
&+\sum_{i\ne j}{\mu^2\over 8\pi^4}\left[\int d^2xd^2y {\langle\Tb(\bar{x})\Tb(\bar{y})\rangle_0\over (z_i-x)(z_i-y)}\right]_{\rm p.s.}
\langle\p^2T(z_i)T(z_j)\rangle+\cdots\\
&={c\over 2z_{12}^4}
+{5c^2\mu^2\over 6\pi^2z_{12}^6\zb_{12}^2}+{\rm div}+{\cal O}(\mu^3)
\end{aligned}
\end{equation}
where we used $\langle T(z)\rangle_0=\langle\Tb(z)\rangle_0=0$ and $[\cdots]_{\rm p.s.}$ denotes a point-splitting regularization similar to the one (implicitly) used in \eqref{2ptZ2nd}.
Once again, div denotes the $1/\epsilon^2$ divergence. 
In the meantime, in a similar way to the computation of \eqref{2ptZ2nd}, the RHS reads
\begin{equation}
\begin{aligned}
\left\langle {c\over 2(Z_1-Z_2)^4}\right\rangle_0& ={c\over 2z_{12}^4}+{5c\mu^2\over 4\pi^4z_{12}^6}
\left\langle\left(\int{d^2x \Tb(\bar{x})\over z_1-x}-\int{d^2y \Tb(\bar{y})\over z_2-y} \right)^2\right\rangle_0+\cdots\\
&={c\over 2z_{12}^4}+{5c^2\mu^2\over 6\pi^2z_{12}^6\zb_{12}^2}+{\rm div}+{\cal O}(\mu^3)\ .
\end{aligned}
\end{equation}
Thus the two agree to second order after renormalization. We note that this second-order result agrees with that of the two-point correlator of $T^{(\mu)}(z_i)$'s in \cite{Kraus:2018xrn, Aharony:2018vux, Hirano:2020ppu} because the Jacobian factor can only start contributing at third order in this case.

\section{Semi-heavy correlators via a simple map}
\label{sec:heavycorrelators}

In the last section, we mostly discussed the practical use of the ${\cal T}^{(0)}[\mathbb{R}_{(0|\mu)}^2]$-theory description of the $T\Tb$-deformed CFT in the study of correlators, but we may have stopped short of offering conceptual perspectives. In this section, we study a semi-classical regime in which one can take full advantage of the fact that the ${\cal T}^{(0)}[\mathbb{R}_{(0|\mu)}^2]$-theory is a CFT which is related to the ${\cal T}^{(\mu)}[\mathbb{R}_{(\mu|0)}^2]$-theory via a simple map \eqref{finiteZmap}. Namely, we focus on the semi-heavy operators, {\it i.e.}, the operators ${\cal O}^{(\mu)}_{\Delta}(z,\zb)\equiv {\cal O}^{(\mu)}_{\Delta, \Delta}(z,\zb)$ of large conformal dimension $\Delta\gg 1$ (more precisely, $\Delta\gg \sqrt{c}$, as we will see below), for which the dynamical coordinate $Z$ becomes approximately classical, {\it i.e.}, a $c$-number coordinate.  As it turns out, we may alternatively interpret this regime as a double-scaling limit, $\mu\to 0$ and $\Delta\to \infty$ keeping $\mu\Delta^2$ fixed finite. From the latter viewpoint, this is a regime in which the $T\Tb$-theory becomes local since the operator size $\sqrt{|\mu|\Delta/\pi}\to 0$.\footnote{This might bear some similarity to the double scaling ('t Hooft-like) limit, $\mu\to 0$ and $c\to\infty$ with $\mu c$ fixed finite, discussed in \cite{Aharony:2018vux}. It might be interesting to see if there is a coherent picture unifying the two limits.}

We claim that for the heavy operators with $\Delta_i\gg \sqrt{c}$, the correlators can be approximated by those of CFT on $\mathbb{R}^2$:
\begin{equation}
\begin{aligned}\label{heavyoperatorcorrelators}
\langle {\cal O}^{(\mu)}_{\Delta_1}(z_1,\bar{z}_1)\cdots {\cal O}^{(\mu)}_{\Delta_n}(z_n,\bar{z}_n)\rangle_0
&=\langle \prod_{i=1}^nJ_i^{\Delta_i}{\cal O}_{\Delta_i}(Z_i){\cal O}_{\Delta_i}(\bar{Z}_i)\rangle_0\\
&\approx\langle\prod_{i=1}^n{\cal O}_{\Delta_i}(Z^{cl}_i)\rangle_0\langle\prod_{i=1}^n{\cal O}_{\Delta_i}(\bar{Z}^{cl}_i)\rangle_0\ ,
\end{aligned}
\end{equation}
where $(Z^{cl}, \bar{Z}^{cl})$ are the $c$-number coordinates and related to $(z,\zb)$ via the map
\begin{align}\label{wZOO}
dz = dZ^{cl} -{\mu\over\pi}\langle\Tb(\bar{Z}^{cl})\rangle_{{\cal O}^n_{\Delta}} d\bar{Z}^{cl}\qquad\mbox{with}\qquad
\langle\Tb(\bar{Z}^{cl})\rangle_{{\cal O}^n_{\Delta}}\equiv \frac{\langle \Tb(\bar{Z}^{cl})\prod_{i=1}^n\bar{\cal O}_{\Delta_i}(\bar{Z}^{cl}_i)\rangle_0}{\langle \prod_{i=1}^n\bar{\cal O}_{\Delta_i}(\bar{Z}^{cl}_i)\rangle_0}\ .
\end{align}
 Since $Z^{cl}$ takes values in the set $\{Z^{cl}_i\}$ $(i=1,\cdots, n)$ in \eqref{heavyoperatorcorrelators}, this definition requires a point-splitting regularization: $Z^{cl}=Z^{cl}_i+\epsilon_i$. Note that $\langle\Tb(\bar{Z}^{cl})\rangle_{{\cal O}_{\Delta}^n}$ is the normalized vev of the stress tensor in the presence of $n$ operator insertions.
In the second line of \eqref{heavyoperatorcorrelators}, the Jacobian factors dropped out since the Jacobian $J_i^{-1}=1-{\mu^2\over\pi^2}T(Z_i)\Tb(\bar{Z}_i)\approx 1$ to leading order at large $\Delta$ as we will see.

\subsection{``Classicalization'' and a factorization property}\label{subsec:factorization2pt}

The key to show the claim \eqref{heavyoperatorcorrelators} is the ``classicalization'' of the dynamical coordinate $Z$ at large $\Delta$: If $Z$ can be replaced by a $c$-number as opposed to an operator, the operators ${\cal O}_{\Delta}(Z)$ become the ordinary CFT operators and the correlators can be computed as those of CFT on $\mathbb{R}^2$. We are now going to show that $Z$ can indeed be replaced by a $c$-number at large $\Delta\gg \sqrt{c}$.

Recall the finite map \eqref{finiteZmap}, {\it i.e.}, the inverse map of \eqref{wZOO}: 
\begin{align}\label{finiteZmapschematic}
Z=z+{\mu\over 2\pi^2}\int d^2x{\Tb^{(\mu)}(x,\bar{x})\over x-z}\equiv z +\delta z[{\cal D}(\mu T), {\cal D}(\mu\Tb)]\ ,
\end{align}
where the notation $\delta z[{\cal D}(\mu T), {\cal D}(\mu \Tb)]$ is to stress that the deformation of the coordinate $\delta z$ is a function of $\mu T$ and $\mu\Tb$ and their derivatives and integrals denoted symbolically by ${\cal D}$.
As discussed in the last section, the basic idea in the computation of correlators is to perform the expansion of the operator
\begin{align}\label{Oexpansion}
{\cal O}_{\Delta}(Z)={\cal O}_{\Delta}(z)+\delta z\partial{\cal O}_{\Delta}(z)+{1\over 2}\delta z^2\partial^2{\cal O}_{\Delta}(z)+\cdots
\end{align}
and $\delta z$ is further expanded in powers of $\mu$ by using the recursion equation \eqref{DeformedTformula}. For example, to second order, the explicit expansion is given in  \eqref{T2ndorder}. 

It is instructive to first consider the simplest case, {\it i.e.}, two-point correlators, to make the argument for proof more concrete and easier to digest. Schematically, they are expanded as
\begin{equation}
\begin{aligned}
\langle {\cal O}^{(\mu)}_{\Delta}(z_1,\bar{z}_1) {\cal O}^{(\mu)}_{\Delta}(z_2,\bar{z}_2)\rangle_0
&=\sum_{i=1,2}\sum_{n,m, p, q} \mu^{n+m}{\cal D}^n{\cal D}^m\partial_{z_i}^p\bar{\partial}_{z_j}^q\langle T(x)^n{\cal O}_{\Delta}(z_1){\cal O}_{\Delta}(z_2)\rangle_0\\
&\hspace{5cm}\times\langle \Tb(\bar{y})^m\bar{\cal O}_{\Delta}(\bar{z}_1)\bar{\cal O}_{\Delta}(\bar{z}_2)\rangle_0\ .
\end{aligned}
\end{equation}
The ``classicalization'' of $Z$ boils down to that of $T^{(\mu)}$, hence of $(T, \Tb)$, as indicated in \eqref{finiteZmapschematic}.
Indeed, the following factorization property holds for $\Delta\gg\sqrt{c}$:
\begin{align}\label{2ptfactorization}
\frac{\langle\prod_{i=1}^n T(x_i){\cal O}_{\Delta}(z_1){\cal O}_{\Delta}(z_2)\rangle_0}{\langle{\cal O}_{\Delta}(z_1){\cal O}_{\Delta}(z_2)\rangle_0}
\approx
\prod_{i=1}^n\frac{\langle T(x_i){\cal O}_{\Delta}(z_1){\cal O}_{\Delta}(z_2)\rangle_0}{\langle{\cal O}_{\Delta}(z_1){\cal O}_{\Delta}(z_2)\rangle_0}\ .
\end{align}
In other words, $T(x)$ can be replaced by a $c$-number, {\it i.e.}, an expectation value in the two-point correlators.
This follows from the conformal Ward-Takahashi (WT) identity,
\begin{equation}
\begin{aligned}
\langle T(x)T(y){\cal O}_{\Delta}(z_1){\cal O}_{\Delta}(z_2)\rangle_0
&={c\over 2(x-y)^4}\langle {\cal O}_{\Delta}(z_1){\cal O}_{\Delta}(z_2)\rangle_0\\
&+\left[{2\over (x-y)^2}+{1\over x-y}\partial_y\right]\langle T(y){\cal O}_{\Delta}(z_1){\cal O}_{\Delta}(z_2)\rangle_0\\
&+\sum_{i=1,2}\left[{\Delta\over (x-z_i)^2}+{1\over x-z_i}\partial_{z_i}\right]\langle T(y){\cal O}_{\Delta}(z_1){\cal O}_{\Delta}(z_2)\rangle_0\ .
\end{aligned}
\end{equation}
At large $\Delta\gg\sqrt{c}$, the third line dominates and we find that
\begin{equation}
\begin{aligned}\label{2ptWTfactorization}
\langle T(x)T(y){\cal O}_{\Delta}(z_1){\cal O}_{\Delta}(z_2)\rangle_0
&\approx\sum_{i=1,2}\left[{\Delta\over (x-z_i)^2}+{1\over x-z_i}\partial_{z_i}\right]\langle T(y){\cal O}_{\Delta}(z_1){\cal O}_{\Delta}(z_2)\rangle_0\ ,
\end{aligned}
\end{equation}
where the three-point correlator on the RHS is given by
\begin{align}
\langle T(x){\cal O}_{\Delta}(z_1){\cal O}_{\Delta}(z_2)\rangle_0 ={\Delta\over (x-z_1)^2(x-z_2)^2z_{12}^{2(\Delta-1)}}
\end{align}
and the leading-order WT identity \eqref{2ptWTfactorization} is of order ${\cal O}(\Delta^2)\gg {\cal O}(c)$.
Explicitly, this yields
\begin{equation}
\begin{aligned}
\langle T(x)T(y){\cal O}_{\Delta}(z_1){\cal O}_{\Delta}(z_2)\rangle_0
&\approx{\left(z_{12}^2\Delta\right)^2\over (x-z_1)^2(x-z_2)^2(y-z_1)^2(y-z_2)^2z_{12}^{2\Delta}}\ .
\end{aligned}
\end{equation}
By iterating the same argument for more insertions of the stress tensor $T(x_i)$, it is straightforward to find that\footnote{As a technical note, the derivative $\partial_{z_i}$ in \eqref{2ptWTfactorization} only acts on the factor $1/z_{12}^{2(\Delta-1)}$ to leading order in $\Delta$. So the factors involving $x_i$ such as $1/(x_i-z_j)^2$ go along the ride in the computation, which makes the proof straightforward.}
\begin{align}
\langle \prod_{i=1}^n T(x_i){\cal O}_{\Delta}(z_1){\cal O}_{\Delta}(z_2)\rangle_0
\approx {1\over z_{12}^{2\Delta}}\prod_{i=1}^n{z_{12}^2\Delta\over (x_i-z_1)^2(x_i-z_2)^2}\ .
\end{align}
This indeed implies the factorization \eqref{2ptfactorization} and the ``classicalization'' of $Z$.

Having shown the factorization property, we wish to close the circle of this argument by connecting it to the claim \eqref{heavyoperatorcorrelators}  in the beginning of this section. 
First, since the dynamical coordinates $Z$'s become $c$-numbers at large $\Delta$ due to ``classicalization'', they can be replaced by $Z^{cl}$ in the correlators. Second, one can show by a power-counting analysis that the Jacobian factors are subleading and thus drop out:
\begin{equation}
\begin{aligned}\label{nptclassicalZ}
\langle \prod_{i=1}^nJ_i^{\Delta}{\cal O}_{\Delta}(Z_i){\cal O}_{\Delta}(\bar{Z}_i)\rangle_0
&\approx\langle\prod_{i=1}^nJ_i(Z^{cl},\bar{Z}^{cl})^{\Delta}{\cal O}_{\Delta}(Z^{cl}_i){\cal O}_{\Delta}(\bar{Z}^{cl}_i)\rangle_0\\
&\approx\langle\prod_{i=1}^n{\cal O}_{\Delta}(Z^{cl}_i){\cal O}_{\Delta}(\bar{Z}^{cl}_i)\rangle_0\ .
\end{aligned}
\end{equation}
We now elaborate on the power-counting analysis to justify the second line. First of all, we are concerned with the double expansion of the correlators in powers of $\mu$ and $\Delta$ and only keep the greatest power of $\Delta$ at every order in the $\mu$ expansion. 
Using the recursion relations in Section \ref{sec:infinitesimal}, one can show that the Jacobian $J=1-{\mu\over\pi}\Theta^{(\mu)}$, as given in \eqref{JacobianTheta}, is a function of $\mu\Delta$ taking the form $J=1-f(\mu\Delta)$ with $f(\mu\Delta)\sim{\cal O}((\mu\Delta)^2)$.\footnote{For concreteness, the trace $\Theta^{(\mu)}$ is given by \eqref{Theta2ndorder} to second order in $\mu$.} Meanwhile, the expansion \eqref{Oexpansion} indicates that the operator ${\cal O}_{\Delta}(Z)$ is a function of $\mu\Delta^2$ to leading order at large $\Delta$ in the $\mu$ expansion since $\delta z= g(\mu\Delta)\sim {\cal O}(\mu\Delta)$ and $\partial_z\sim{\cal O}(\Delta)$. So at every order in $\mu$, we see that the Jacobian factors $J^{\Delta}$ can only yield subleading contributions in powers of $\Delta$ and thus can be dropped. As an important point, implied in this argument is that the correlators are functions of $\mu\Delta^2$ to leading order at large $\Delta$ in the $\mu$ expansion. This, in particular, suggests the aforementioned double scaling limit $\mu\to 0$ and $\Delta\to\infty$ with $\mu\Delta^2$ fixed finite.
Finally, in the coordinate transformation \eqref{wZOO} between $(z, \zb)$ and $(Z, \bar{Z})$, the stress tensor $T(Z)$ (or its complex conjugate) gets replaced by a $c$-number, {\it i.e.}, the expectation value $\langle T(Z)\rangle_{{\cal O}_{\Delta}^n}$ in the presence of $n$ operators ${\cal O}_{\Delta_i}(Z_i)$ ($i=1,\cdots,n$) in this semiclassical regime. 

\subsubsection{Generalization to higher-point correlators}\label{subsec:higher}

We are now going to generalize the proof of the factorization \eqref{2ptfactorization} to higher-point correlators. By a similar argument, the conformal WT identity approximates to
\begin{equation}
\begin{aligned}\label{higherptfactorization}
\hspace{-.48cm}
\langle T(x)T(y)\prod_{i=1}^n{\cal O}_{\Delta_i}(z_i)\rangle_0
&\approx\sum_{i=1}^n\left[{\Delta_i\over (x-z_i)^2}+{\partial_{i}\over x-z_i}\right]\!\langle T(y)\prod_{i=1}^n{\cal O}_{\Delta_i}(z_i)\rangle_0\\
&\approx\sum_{i,j=1}^n\!\left[{\Delta_i\over (x-z_i)^2}+{\partial_{i}\over x-z_i}\right]\!\left[{\Delta_j\over (y-z_j)^2}+{\partial_{j}\over y-z_j}\right]\!
\langle \prod_{i=1}^n{\cal O}_{\Delta_i}(z_i)\rangle_0
\end{aligned}
\end{equation}
to leading order in  large $\Delta_i\gg \sqrt{c}$. This is a quantity of order $\Delta^2$ and the approximate equality $\approx$ means an equality up to ${\cal O}(\Delta)$ corrections.
So the factorization 
\begin{equation}
\begin{aligned}\label{2Tfactor}
\frac{\langle T(x)T(y)\prod_{i=1}^n{\cal O}_{\Delta_i}(z_i)\rangle_0}{\langle \prod_{i=1}^n{\cal O}_{\Delta_i}(z_i)\rangle_0}
&\approx\frac{\langle T(x)\prod_{i=1}^n{\cal O}_{\Delta_i}(z_i)\rangle_0}{\langle \prod_{i=1}^n{\cal O}_{\Delta_i}(z_i)\rangle_0}
\frac{\langle T(y)\prod_{i=1}^n{\cal O}_{\Delta_i}(z_i)\rangle_0}{\langle \prod_{i=1}^n{\cal O}_{\Delta_i}(z_i)\rangle_0}
\end{aligned}
\end{equation}
requires that
\begin{align}\label{factorcond}
\hspace{-.5cm}
\sum_{i,j=1}^n{\langle \prod_{i=1}^n{\cal O}_{\Delta_i}(z_i)\rangle_0\partial_{i}\partial_{j}\langle \prod_{i=1}^n{\cal O}_{\Delta_i}(z_i)\rangle_0\over (x-z_i)(y-z_j)}
&\approx \sum_{i,j=1}^n{\partial_{i}\langle \prod_{i=1}^n{\cal O}_{\Delta_i}(z_i)\rangle_0\partial_{j}\langle\prod_{i=1}^n{\cal O}_{\Delta_i}(z_i)\rangle_0\over (x-z_i)(y-z_j)}
\end{align}
which can be rewritten as
\begin{align}
\sum_{i,j=1}^n\frac{\partial_{i}\partial_{j}\ln\langle\prod_{i=1}^n{\cal O}_{\Delta_i}(z_i)\rangle_0}{(x-z_i)(y-z_j)}={\cal O}(\Delta)\ll {\cal O}(\Delta^2)\ .
\end{align}
This is indeed the case since $\ln\langle\prod_{i=1}^n{\cal O}_{\Delta_i}(z_i)\rangle= {\cal O}(\Delta)$. So the factorization property generalizes to higher-point correlators of the heavy operators ${\cal O}_{\Delta_i}(z)$ with $\Delta_i\gg\sqrt{c}$.
To generalize this proof to more insertions of the stress tensor $T(x_i)$, note that each insertion adds a factor of the differential operator in square bracket in \eqref{higherptfactorization} to the WT identity. With the repeated use of \eqref{factorcond} for every additional insertion of $T(x)$, the factorization for two insertions of $T$'s \eqref{2Tfactor} generalizes to 
\begin{equation}
\begin{aligned}\label{manyTfactor}
\frac{\langle \prod_aT(x_a)\prod_{i=1}^n{\cal O}_{\Delta_i}(z_i)\rangle_0}{\langle \prod_{i=1}^n{\cal O}_{\Delta_i}(z_i)\rangle_0}
&\approx\prod_a\frac{\langle T(x_a)\prod_{i=1}^n{\cal O}_{\Delta_i}(z_i)\rangle_0}{\langle \prod_{i=1}^n{\cal O}_{\Delta_i}(z_i)\rangle_0}\ .
\end{aligned}
\end{equation}
Combining with the power-counting argument below \eqref{nptclassicalZ}, this proves the claim \eqref{heavyoperatorcorrelators} in the beginning of this section.

\subsection{The two-point correlators}\label{subsec:2ptcorrelators}

As a check and further evidence for the claim \eqref{heavyoperatorcorrelators}, we show in detail how to compute the two-point correlators, reproducing the known results in the literature. For the two-point correlators, the formula \eqref{heavyoperatorcorrelators} reads
\begin{equation}
\begin{aligned}\label{heavyoperator2pt}
\langle {\cal O}^{(\mu)}_{\Delta}(z_1,\bar{z}_1) {\cal O}^{(\mu)}_{\Delta}(z_2,\bar{z}_2)\rangle_0
&\approx \frac{1}{\left|Z^{cl}_1 - Z^{cl}_2\right|^{4\Delta}}\ ,
\end{aligned}
\end{equation}
where the coordinate transformation between $(z,\zb)$ and $(Z, \bar{Z})$ is given by
\begin{equation}
\begin{aligned}\label{zZ2ptmap}
z_i&=\int\left(dZ^{cl}_i-{4a^2\mu\Delta d\bar{Z}_i^{cl}\over \pi(\bar{Z}_i^{cl}-a)^2(\bar{Z}_i^{cl}+a)^2}\right)\\
&=Z_i^{cl}+{\mu\Delta \over \pi}\left({1\over \bar{Z}_i^{cl}-a}+{1\over \bar{Z}_i^{cl}+a}-{1\over a}\ln{\bar{Z}_i^{cl}+a\over \bar{Z}_i^{cl}-a}\right)
\end{aligned}
\end{equation}
for the operator insertions at $Z^{cl}_1=\bar{Z}^{cl}_1=-a$ and $Z^{cl}_2=\bar{Z}^{cl}_2=a$.\footnote{This map has already appeared in \eqref{z_vs_Z_2pt} and some aspects of it was discussed in Section \ref{sec:TTbardeformedspace}.} As commented below \eqref{wZOO}, this requires a point-splitting regularization: 
$Z^{cl}_1\to -a - \epsilon$ and $Z^{cl}_2\to a+\epsilon$. The two-point correlator is that of CFT and as simple as
\begin{equation}
\begin{aligned}\label{2pta}
\langle {\cal O}^{(\mu)}_{\Delta}(z_1,\bar{z}_1) {\cal O}^{(\mu)}_{\Delta}(z_2,\bar{z}_2)\rangle_0
&\approx \frac{1}{(2a)^{4\Delta}}\ .
\end{aligned}
\end{equation}
The $\mu$-dependent information of the $T\Tb$-deformation is all encoded in the map \eqref{zZ2ptmap}. To find the $T\Tb$-deformed correlator, we need to express \eqref{2pta} in terms of the flat coordinates $(z_1, \zb_1)=(-b, -b)$ and $(z_2, \zb_2)=(b, b)$. Inverting the map \eqref{zZ2ptmap}, we find 
\begin{equation}
\begin{aligned}\label{aofb}
a&=b-\epsilon-{\mu\Delta \over \pi}\left({1\over \epsilon}+{1\over 2b-\epsilon}-{1\over b-\epsilon}\ln{2b-\epsilon\over \epsilon}\right)\\
&\quad-{\mu^2\Delta^2\over\pi^2(2b-\epsilon)^2}\left({1\over \epsilon}+{1\over 2b-\epsilon}-{1\over b-\epsilon}\ln{2b-\epsilon\over \epsilon}\right)\\
&\quad+{\mu^2\Delta^2\over\pi^2(b-\epsilon)^2}\left({1\over \epsilon}+{1\over 2b-\epsilon}-{1\over b-\epsilon}\ln{2b-\epsilon\over \epsilon}\right)\ln{2b-\epsilon\over \epsilon}\\
&\quad-{\mu^2\Delta^2\over\pi^2(b-\epsilon)(2b-\epsilon)}\left({1\over \epsilon}+{1\over 2b-\epsilon}-{1\over b-\epsilon}\ln{2b-\epsilon\over \epsilon}\right)
+{\cal O}(\mu^3)\ .
\end{aligned}
\end{equation}
For reference, we presented the result to second order in the $\mu$ expansion. This is already cumbersome and it seems hopeless to find the all-order expression. However, fortunately, we actually need only the first-order result (first line) to compute the all-order two-point correlators at large $\Delta$.\footnote{The first-order correction agrees with \eqref{2pt1storder} including the regularized divergences.} For example, the second-order contributions in \eqref{aofb} can only yield ${\cal O}(\mu^2\Delta^3)\ll {\cal O}((\mu\Delta^2)^2)$ to second-order in $\mu$ in expanding \eqref{2pta}.
More generally, one can easily convince oneself that the leading $\Delta$ contributions to the correlators only come from the first-order expansion.
Denoting $a=(b-\epsilon)(1 -\delta b) +{\cal O}(\mu^2)$ where $-(b-\epsilon)\delta b$ is the first-order correction in \eqref{aofb}, the two-point correlators are then given by 
\begin{equation}
\begin{aligned}
\langle {\cal O}^{(\mu)}_{\Delta}(-b,-b) {\cal O}^{(\mu)}_{\Delta}(b,b)\rangle_0
&\approx{1\over (2(b-\epsilon))^{4\Delta}}e^{4\Delta\delta b}
\end{aligned}
\end{equation}
to leading order in $\Delta$, where the renormalization of the $1/\epsilon^{n}$ divergences is assumed.\footnote{The power divergences can be renormalized by redefining the operators as
\begin{align}
\widetilde{\cal O}_{\Delta}^{(\mu)}(\pm b,\pm b)\equiv e^{-{2\mu\Delta^2\over\pi b\epsilon}}{\cal O}_{\Delta}^{(\mu)}(\pm b,\pm b)\ .
\end{align}} In particular, the leading-log contribution reads
\begin{align}
\langle {\cal O}^{(\mu)}_{\Delta}(-b,-b) {\cal O}^{(\mu)}_{\Delta}(b,b)\rangle_0
\approx\sum_{n=0}^{\infty}(-1)^n{(16\mu\Delta^2)^n\over n!\pi^n}{\ln^n(2b/\epsilon)\over (2b)^{4\Delta+2n}}+\cdots\ ,
\end{align}
where the dots are the terms subleading in powers of $\ln(2b/\epsilon)$. This perfectly agrees, to leading order in $\Delta$, with the known all-order leading-log contributions to the two-point correlators \cite{Cardy:2019qao}:
\begin{equation}
\begin{aligned}\label{Cardy2ptallorder}
\langle {\cal O}^{(\mu)}_{\Delta}(\vec{x}) {\cal O}^{(\mu)}_{\Delta}(0)\rangle_0
&={\Gamma\left(1-2\Delta\right)\over \pi 2^{4\Delta}\Gamma\left(2\Delta\right)}\int_{-\infty}^{\infty}d^2\vec{k} e^{i\vec{k}\cdot\vec{x}}
k^{2(2\Delta-1)}e^{-{\mu\over 2\pi} k^2\ln (k^2\epsilon^2)}+\cdots\\
&=\sum_{n=0}^{\infty}(-1)^n2^{2n}{\mu^n\over n!\pi^n}\prod_{k=0}^{n-1}(2\Delta+k)^2{\ln^n(|x|/\epsilon)\over |x|^{4\Delta+2n}}+\cdots\ .
\end{aligned}
\end{equation}
This check serves as evidence for the claim \eqref{heavyoperatorcorrelators}.
It should, however, be noted that the subleading $\Delta$-corrections in \eqref{heavyoperator2pt} do not agree with those in \eqref{Cardy2ptallorder} as they are not supposed to.
We have not found a way to systematically include the $1/\Delta$ corrections in this semiclassical approximation. 
However, as discussed earlier in Section \ref{sec:correlators}, we can, in principle, compute the exact correlators from \eqref{deformedcorrelators} and \eqref{primaryoperator_map} order by order in the $\mu$ expansion.

\section{Comments on $T\Tb$-deformed correlators from cutoff AdS}\label{sec:2ptcutoffAdS}

As alluded to in the beginning of Section \ref{sec:TTbardeformedspace}, the $T\Tb$-deformed CFT is conjectured to be dual to quantum gravity on an $AdS_3$ space \cite{Maldacena:1997re} with a finite radial cutoff \cite{McGough:2016lol}. 
One of the outstanding issues in this proposal is the inclusion of bulk matter. Namely, it lacks a working GKPW dictionary \cite{Gubser:1998bc, Witten:1998qj} and it has not been understood how to compute the matter correlators in the cutoff AdS as first discussed in   \cite{Kraus:2018xrn}. In other words, the naive form of the cutoff AdS proposal only works for pure gravity without matter, which may be foreseen by the fact that the flow equation \eqref{floweqn} can be identified with the (purely) radial component of the Einstein equations only in the absence of matter  \cite{Kraus:2018xrn, Caputa:2020lpa}. 

\begin{figure}[h!]
\centering
\includegraphics[scale=0.4]{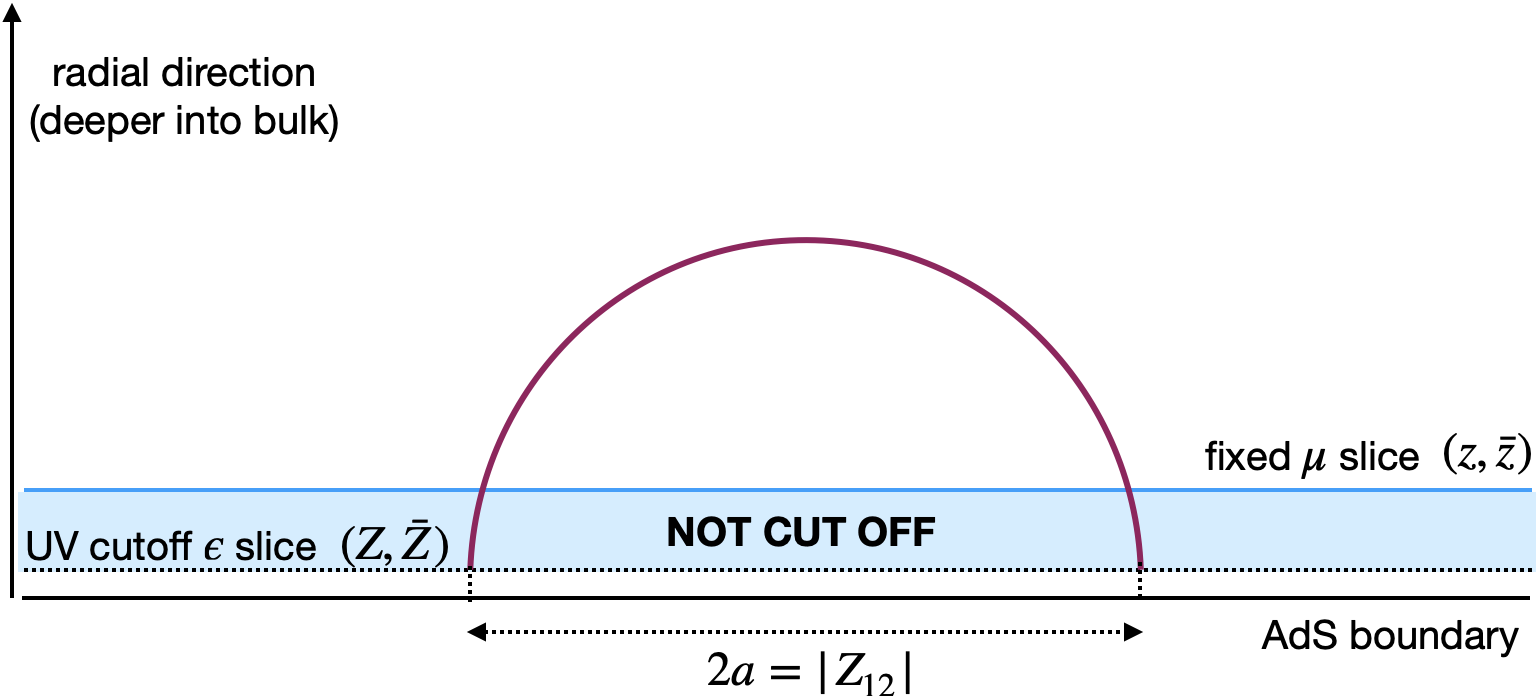}
\caption{\sl An image of the holographic dictionary for the two-point correlator \eqref{2pta} in the geodesic approximation for illustration: The solid purple curve (semi-circle) represents the geodesic between the two points $(Z_1,\bar{Z}_1)$ and $(Z_2,\bar{Z}_2)$ on the UV cutoff surface. The parts of the geodesic in the strip (indicated in light blue) between the constant $\mu$ and $\epsilon$ slices do contribute to the two-point correlator. So the constant $\mu$ slice is not a cutoff surface. Instead, the distance $2a=|Z_{12}|$ is remeasured in terms of $2b=|z_{12}|$ in the flat coordinates $(z,\zb)$ on the fixed $\mu$ slice determined by \eqref{zZ2ptmap}.}
\label{fig:AdSCFT}
\end{figure}

Even though it is out of scope of this paper to fully address this issue, we would like to discuss an implication of our findings in Section \ref{sec:heavycorrelators} and offer an idea which might partially solve the issue of matter correlators in the cutoff AdS proposal. The formula \eqref{heavyoperatorcorrelators} applies, in particular, to the heavy operators, {\it i.e.}, the operators of dimension $\Delta\sim{\cal O}(c)\gg \sqrt{c}$ at large $c$. In this case, the Ba\~nados metric \eqref{3Dbulkmetric} describes the $AdS_3$ space with conical defects \cite{Martinec:1998wm, Matschull:1998rv, Balasubramanian:1999zv}. The CFT correlators with all heavy operators then correspond to the exponential of (minus) the on-shell action on these conical defect geometries \cite{Chang:2016ftb}. To obtain the $T\Tb$-deformed correlators, we simply perform the coordinate transformation \eqref{wZOO} from $(Z^{cl}_i, \bar{Z}^{cl}_i)$ to the flat coordinates $(z_i, \zb_i)$ on a \emph{fixed} radial slice (corresponding to the $T\Tb$-coupling $\mu$) as opposed to the variable radial slice which would bring the Ba\~nados metric to the Poincar\'e $AdS_3$ \cite{Roberts:2012aq}.
In fact, the argument in Section  \ref{sec:heavycorrelators} suggests that this works similarly for a class of light operators of dimension $\sqrt{c}\ll\Delta\ll c$. For these operators, the probe approximation suffices. So the standard GKPW dictionary followed by the coordinate transformation \eqref{wZOO} gives the $T\Tb$-deformed correlators at large $\Delta$.
To be clear, there is no actual radial cutoff in this idea and what is described above is the standard AdS/CFT except that the correlators are remeasured in the coordinate distance, determined by \eqref{wZOO}, at a fixed radial slice as illustrated in Figure \ref{fig:AdSCFT}.

Of course, the issue still remains as this only works for all semi-heavy correlators at leading order in $\Delta$. We have not understood how, in principle, to holographically incorporate the subleading corrections in $\Delta$ even in the large $c$ limit.

\section{Discussions}\label{sec:Discussions}

We studied the map among different descriptions of the $T\Tb$-deformed conformal field theory, labeled as ${\cal T}^{(\lambda_1)}[\mathbb{R}^2_{(\lambda_1|\lambda_2)}]$ with $\lambda_1+\lambda_2=\mu$, which follows from the interpretation of the $T\Tb$ deformation as a (dynamical) coordinate transformation.
This perspective is very useful, both technically and conceptually, for the study of the $T\Tb$-deformed CFT\@. 
Technically, it leads to systematic and straightforward computations of the $T\Tb$-deformation of the stress tensor, operators, and their correlators, while rederiving known results in the literature. Conceptually, it offers a novel type of QFT which lives on a dynamical space, reminiscent of the theory of gravity, 
and as discussed in Section \ref{sec:TTbardeformedspace}, it further gives a rather direct way to study the short-distance properties of the $T\Tb$ deformation.

Among these descriptions, particularly useful and appealing is the ${\cal T}^{(0)}[\mathbb{R}^2_{(0|\mu)}]$-theory, that is, the undeformed CFT on the $T\Tb$-deformed $\mathbb{R}^2$. As discussed in Section \ref{sec:heavycorrelators}, this description as a CFT shows its full advantage in the semiclassical regime where the conformal dimension $\Delta$ of the operators are so large as $\Delta\gg\sqrt{c}$, and it gives an intuitive and simple way to compute the $T\Tb$-deformed correlators from the CFT ones via dynamical coordinate transformations.
As discussed in Section \ref{sec:2ptcutoffAdS}, the last point has implications in the holographic dual description, which points towards a working dictionary for a class of matter correlators in the cutoff AdS picture. However, it fell short of offering a comprehensive solution to the issue of the cutoff AdS proposal in the presence of matter.
The challenge with this idea is if there is an intuitive and systematic way to include the $1/\Delta$ corrections in the semiclassical framework in Section \ref{sec:heavycorrelators} and if it admits a natural holographic translation.

As a final point, one of the underlying themes of this paper was to study conformal field theory on $T\Tb$-deformed space, ${\cal T}^{(0)}[\mathbb{R}^2_{(0|\mu)}]$.
However, in contrast to ordinary CFT, the coupling between the holomorphic and anti-holomorphic sectors played an essential role as a description of the $T\Tb$-deformed CFT\@. 
It is desirable to have a conceptual understanding of this point than merely to accept it as a technical fact. One natural idea is to understand this coupling as a contact interaction. Namely, even in ordinary CFT, there is a coupling between the two sectors in the form of contact terms. For example, the contact term of the holomorphic and anti-holomorphic components of the stress tensor is given by $\langle T(z_1)\Tb(\zb_2)\rangle= -{\pi c\over 6}\p\pb\delta^2(z_{12})$.
However, as discussed in Section \ref{sec:TTbardeformedspace}, the $T\Tb$-deformed space exhibits some degree of non-locality. So the notion of $\delta$-function must be suitably generalized. 
As an illustration, we take the two-point correlator of $T$ and $\Tb$ as an example. Using \eqref{T2ndorder}, one finds that
\begin{equation}
\begin{aligned}\label{2ndTZ1TZ2}
\langle T(Z_1)\Tb(\bar{Z}_2)\rangle&=-{\pi c\over 6}\p\pb\delta^2(z_{12})+{\mu^2c^2\over 4\pi^2z_{12}^4\zb_{12}^4}+{\rm contacts}+{\rm div}+{\cal O}(\mu^3)\ ,
\end{aligned}
\end{equation}
where contacts denotes the contact terms of order ${\cal O}(\mu^2)$.
We focus our discussion on the second non-contact term and propose a generalization of the $\delta$-function that reproduces it. 
The idea is to start with the identity
\begin{align}
\delta^2(z)={1\over 2\pi}\pb{1\over z}={1\over 4\pi}\left[\pb{1\over z}+\p{1\over \zb}\right]
\end{align}
and define the generalized $\delta$-function by
\begin{align}
{\Delta}^2(Z)&\equiv {1\over 4\pi}\left[{\p\over\p\bar{Z}}{1\over Z}+{\p\over\p Z}{1\over \bar{Z}}\right]
={1\over 4\pi}\biggl[\pb{1\over Z}+\p{1\over \bar{Z}}
-{{\mu\over\pi}T^{(\mu)}\over 1-{\mu\over\pi}\Theta^{(\mu)}}\p{1\over Z}
-{{\mu\over\pi}\Tb^{(\mu)}\over 1-{\mu\over\pi}\Theta^{(\mu)}}\pb{1\over \bar{Z}}\biggr]\ .
\end{align}
So we propose that 
\begin{align}
\langle T(Z_1)\Tb(\bar{Z}_2)\rangle=-{\pi c\over 6}\left\langle{\p^2\over \p Z_1\p\bar{Z}_1}\Delta^2(Z_{12})\right\rangle_0\ .
\end{align}
Using the finite map \eqref{finiteZmap} and expanding $(Z_i, \bar{Z}_i)$ in powers of $\mu$, the first nontrivial (non-divergent) non-contact term is calculated as
\begin{equation}
\begin{aligned}
\langle T(Z_1)\Tb(\bar{Z}_2)\rangle&=-{\pi c\over 6}\p\pb\delta^2(z_{12})-{\pi c\over 6}\p_1\pb_1
\biggl[-{\mu^2\over 4\pi^4z_{12}^2}
\int {d^2x\langle\Tb(\zb_1)\Tb(\bar{x})\rangle_0\over (z_1-x)(z_2-x)}\biggr]+{\rm c.c.}+\cdots\\
&=-{\pi c\over 6}\p\pb\delta^2(z_{12})+{\mu^2c^2\over 4\pi^2 z_{12}^4\zb_{12}^4}+{\rm div}+\cdots.
\end{aligned}
\end{equation}
This indeed reproduces the second non-contact term in  \eqref{2ndTZ1TZ2}.
On the first pass, this idea seems to work. However, it remains to be seen if it works in generality. If it does, we would have a better understanding of the $T\Tb$-deformed CFT in terms of the CFT on the $T\Tb$-deformed space.


\section*{Acknowledgments}

SH would like to thank Robert de Mello Koch for discussions and Ivonne Zavala for suggesting him an important reference. 
He also would like to thank the Departments of Mathematics and Physics at Nagoya University for hospitality and  the Isaac Newton Institute for Mathematical Sciences, Cambridge, for support and hospitality during the programme, Black holes: bridges between number theory and holographic quantum information, where work on this paper was partially undertaken. This work was supported in part by EPSRC grant no EP/R014604/1 and
the work of SH was supported in part by  the National Natural Science Foundation of China under Grant No.12147219.
The work of MS
was supported in part by MEXT KAKENHI Grant Numbers 21K03552 and
21H0518.

\appendix


\section{Equivalence between \eqref{infinitesimalZmap} and line integral representation }
\label{app:lineintegral}

\begin{figure}[h]
 \begin{center}
 \begin{tikzpicture}[scale=1.5]
  \draw (-0.8,0.85) -- +(0.3,0) -- +(0.3,0.3);
 \node () at (-0.65,1) {$x$};
 \draw (0,0) -- (1,1);
 \draw [thick,black,domain=70:380]  plot ({0.15*cos(\x)}, {0.15*sin(\x)});
 \draw [thick,black,domain=70:380]  plot ({1-0.15*cos(\x)}, {1-0.15*sin(\x)});
 \draw[thick] (0.1-0.05,0.1+0.05) -- (0.9-0.05,0.9+0.05);
 \draw[thick] (0.1+0.05,0.1-0.05) -- (0.9+0.05,0.9-0.05);
 \draw[thick,-latex] (0.5-0.05,0.5+0.05) -- +(0.1,0.1) node [left,xshift=-3] {$C$};
 \draw[thick,-latex] (0.5+0.05,0.5-0.05) -- +(-0.1,-0.1);
 \draw[fill=black] (0,0) circle (0.05) node [right,xshift=3,yshift=-3] {$X$};
 \draw[fill=black] (1,1) circle (0.05) node [right,xshift=3] {$z$};
 \end{tikzpicture}
  \caption{\sl Contour $C$\label{fig:contour_C}}
 \end{center}
\end{figure}
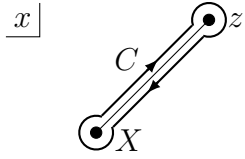

In this appendix, we give more details about the expression \eqref{infinitesimalZmap} for the infinitesimal map:
\begin{align}\label{app:infinitesimalZmap}
Z^{(\mu|\delta\mu)}(z,\zb)=z+{\delta\mu\over 2\pi^2}\int_{\mathbb{R}^2}d^2x\,\bar{T}^{(\mu)}(x,\bar{x})\left({1\over z-x}-{1\over X-x}\right)\ ,
\end{align}
where we modified \eqref{infinitesimalZmap} to include a reference point
$(X,\Xb)$; the original expression \eqref{infinitesimalZmap} corresponds to
the choice $X=\infty$.  In an earlier work \cite{Cardy:2019qao}, a
different expression was given in the line integral form
\begin{align}
\label{CardyLineInt}
Z^{(\mu|\delta\mu)}(z,\zb)=z+{\delta\mu\over\pi}
\int^{(z,\zb)}_{(X,\bar{X})} \left(d\bar{x}\,\bar{T}^{(\mu)}(x, \bar{x})-dx \,\Theta^{(\mu)}(x,\bar{x})\right)
\ .
\end{align}
To show that the two expressions are identical,
note that
\begin{align}
 &dx\wedge d\xb \,\,\Tb^{(\mu)}(x,\xb)\left({1\over z-x}-{1\over X-x}\right)
 \notag\\
 &=-d\left[\log\biggl({x-z\over x-X}\biggr)\, \Tb^{(\mu)}(x,\xb)d\xb\right]
 +\log\left({x-z\over x-X}\right)\,\p\Tb^{(\mu)}(x,\xb)\,dx\wedge d\xb
 \notag\\
 &=-d\left[\log\biggl({x-z\over x-X}\biggr)\left(\Tb^{(\mu)}(x,\xb)d\xb-\Theta^{(\mu)}(x,\xb)\,dx\right)\right],
\end{align}
where in the last equality we used the conservation law, $\p\Tb^{(\mu)}+\pb\Theta^{(\mu)}=0$.
Therefore, we can rewrite the second term in
\eqref{app:infinitesimalZmap}, upon using $d^2x=i dx\wedge d\xb$, as
\begin{align}
 &-{i\delta \mu\over 2\pi^2}\int _{\mathbb{R}^2}
 d\left[\log\biggl({x-z\over x-X}\biggr)\left(\Tb^{(\mu)}(x,\xb)d\xb-\Theta^{(\mu)}(x,\xb)\,dx\right)\right]\notag\\
 &\qquad\qquad
 =-{i\delta \mu\over 2\pi^2}\oint _C
 \log\biggl({x-z\over x-X}\biggr)\left(\Tb^{(\mu)}(x,\xb)d\xb-\Theta^{(\mu)}(x,\xb)\,dx\right),
\end{align}
where $C$ is the closed contour that goes clockwise around the path
connecting $(z,\zb)$ and $(X,\Xb)$ (see Figure~\ref{fig:contour_C}).
Considering the phase coming from the log (which turns the
closed-contour integral $\int_C$ into the line integral $2\pi
i\int_{(X,\Xb)}^{(z,\zb)}$), this reproduces the second term of~\eqref{CardyLineInt}.

%
%
%
%
%


\section{The stress tensor deformation to second order }
\label{app:secondorder}

By solving the recursion relations \eqref{DeformedTformula} and its complex conjugate, the deformation of the stress tensor is calculated to second order as
\begin{equation}
\begin{aligned}\label{T2ndorder}
T^{(\mu)}&=T+{\mu\over 2\pi^2}\int d^2x {\bar{T}(\bar{x})\partial T(z)\over z-x}+{\mu^2\over \pi^2}\left[T(z)^2\bar{T}(\bar{z})\right]_{\rm p.s.}\\
&+{\mu^2\over 8\pi^4}\left(\int_{\mathbb{R}^2}d^2x {\Tb(\bar{x})\over z-x}\right)_{\rm p.s.}^2\partial^2T(z)
+{\mu^2\over 2\pi^3}\int_{\mathbb{R}^2}d^2x {T(x)\Tb(\zb)\partial T(z)\over \zb-\bar{x}}\ ,
\end{aligned}
\end{equation}
where the subscript p.s. stands for the point-splitting regularization. 
This can be rewritten more concisely as
\begin{equation}
\begin{aligned}\label{T2ndorderV2}
T^{(\mu)}&=T+{\mu\over 2\pi^2}\partial_z\int d^2x {\bar{T}(\bar{x}) T(z)\over z-x}\\
&+{\mu^2\over 8\pi^4}\partial^2_z\left[\left(\int_{\mathbb{R}^2}d^2x {\Tb(\bar{x})\over z-x}\right)_{\rm p.s.}^2T(z)\right]
+{\mu^2\over 2\pi^3}\partial_z\int_{\mathbb{R}^2}d^2x {T(x)\Tb(\zb)T(z)\over \zb-\bar{x}}\ .
\end{aligned}
\end{equation}
Then, using \eqref{ThetabyT}, the trace is found to be
\begin{equation}
\begin{aligned}\label{Theta2ndorder}
\Theta^{(\mu)}&=-{\mu\over\pi}T\bar{T}-{\mu^2\over 2\pi^3}\int d^2x{\left[T(x)^2\bar{\partial}\Tb(\bar{x})\right]_{\rm p.s.}\over\zb-\bar{x}}
-{\mu^2\over 4\pi^4}\int d^2x\int d^2y{\Tb(\bar{y})\Tb(\bar{x})\partial^2T(x)\over(\zb-\bar{x})(x-y)}\\
&-{\mu^2\over 4\pi^4}\int d^2x\int d^2y{T(y)\bar{\partial}\Tb(\bar{x})\partial T(x)\over (\zb-\bar{x})(\bar{x}-\bar{y})}
\end{aligned}
\end{equation}
to second order.


\section{Holographic derivation of the stress tensor map}
\label{app:mapfromgravity}

We review the derivation of the stress tensor map \eqref{TmuTZ} by using a holographic argument \cite{Guica:2019nzm}.
As commented in the beginning of Section \ref{sec:TTbardeformedspace}, the $T\Tb$-deformed space $\mathbb{R}^2_{(0|\mu)}$ is conjectured to be identified with the cut-off boundary of the Ba\~nados space \cite{Banados:1998gg}:
\begin{equation}
\begin{aligned}
ds^2&={dz^2\over z^2}+{\pi\over z^2}
\left(dZ-{z^2\over \pi}\Tb(\bar{Z})d\bar{Z}\right)
\left(d\bar{Z}-{z^2\over \pi}T(Z)dZ\right)
\equiv {dz^2\over z^2}+\gamma_{ab}dZ^adZ^b\ ,
\end{aligned}
\end{equation}
where the radial coordinate $z$ is related to the $T\Tb$ coupling via $z=\sqrt{\mu}$ and we set the unit $4G=1$.\footnote{In this appendix, with an abuse of notation, we use $z$ to denote the radial coordinate of the $AdS_3$ space. It should not be confused with the complex coordinate $z$ in earlier sections. The coordinate $w$, which appears later in this appendix, corresponds to the complex coordinate $z$ for the undeformed $\mathbb{R}^2$.}
In the AdS holography, the boundary stress tensor is identified with the Brown-York tensor \cite{Balasubramanian:1999re} and given by
\begin{align}
T_{ab}=\Theta_{ab}-\gamma_{ab}\Theta+\gamma_{ab}
\qquad\quad\mbox{with}\qquad\quad
\Theta=\gamma^{ab}\Theta_{ab}\ ,
\end{align}
where $\Theta_{ab}$ is the extrinsic curvature on the boundary surface.
The unit normal vector is $n_a=(1/z, 0, 0)$ and so the extrinsic curvature is calculated as
\begin{align}
\Theta_{ab}=-{1\over 2}(\nabla_an_b+\nabla_bn_a)=\Gamma^z_{\mbox{  }ab}n_z
\end{align}
which yields
\begin{align}
\Theta_{Z\bar{Z}}={\pi\over 2z^2}\left(1-{z^4\over\pi^2}T(Z)\Tb(\bar{Z})\right)\ ,\qquad
\Theta_{ZZ}=\Theta_{\bar{Z}\bar{Z}}=0\ .
\end{align}
The boundary stress tensor indeed reads
\begin{align}
T_{ZZ}&=(1-\Theta)\gamma_{ZZ}\quad\xrightarrow{z\to 0}\quad T(Z)\ ,\\
T_{\bar{Z}\bar{Z}}&=(1-\Theta)\gamma_{\bar{Z}\bar{Z}}\quad\xrightarrow{z\to 0}\quad \Tb(\bar{Z})\ ,\\
T_{Z\bar{Z}}&=\Theta_{Z\bar{Z}}+\gamma_{Z\bar{Z}}(1-\Theta)\quad\xrightarrow{z\to 0} \quad 0\ .
\end{align}
Now, with the $T\Tb$-deformation, the stress tensor is conjectured to be identified with the Brown-York tensor at a finite $z$ \cite{McGough:2016lol} with respect to the $(w,\bar{w})$ coordinates in terms of which the boundary metric is manifestly flat, $ds_{\rm bdy}^2=dwd\bar{w}$. The coordinate transformation is given by 
\begin{align}
\begin{pmatrix}
dZ\\
d\bar{Z}
\end{pmatrix}
={1\over 1-{\mu^2\over\pi^2}T\Tb}
\begin{pmatrix}
1 & {\mu\over\pi}\Tb \\
{\mu\over\pi}T & 1
\end{pmatrix}
\begin{pmatrix}
dw\\
d\bar{w}
\end{pmatrix}\ .
\end{align}
This yields
\begin{align}
T^{(\mu)}&\equiv T_{ww}={1\over \left(1-{\mu^2\over\pi^2}T\Tb\right)^2}
\left(T_{ZZ}+{2\mu\over\pi}TT_{Z\bar{Z}}+{\mu^2\over\pi^2}T^2T_{\bar{Z}\bar{Z}}\right)={T\over 1-{\mu^2\over\pi^2}T\Tb}\ ,
\end{align}
where the Brown-York tensor components at a finite $z=\sqrt{\mu}$ are given by
\begin{align}\label{TZtensor}
T_{ZZ}&={1+3{\mu^2\over\pi^2}T\Tb\over 1-{\mu^2\over\pi^2}T\Tb}T\ ,\quad
\Tb_{\bar{Z}\bar{Z}}={1+3{\mu^2\over\pi^2}T\Tb\over 1-{\mu^2\over\pi^2}T\Tb}\Tb\ ,\quad
T_{Z\bar{Z}}=-{\mu\over\pi}{3T\Tb+{\mu^2\over\pi^2}(T\Tb)^2 \over 1-{\mu^2\over\pi^2}T\Tb}\ .
\end{align}
Note that at a finite radial position, these components form a tensor whereas $(T(Z), \Tb(\bar{Z}), 0)$ do not.
Similarly, the trace can be calculated as
\begin{equation}
\begin{aligned}
\Theta^{(\mu)}\equiv T_{w\bar{w}}&={1\over \left(1-{\mu^2\over\pi^2}T\Tb\right)^2}
\left({\mu\over\pi}\Tb T_{ZZ}+\left(1+{\mu^2\over\pi^2}T\Tb\right)T_{Z\bar{Z}}+{\mu\over\pi}TT_{\bar{Z}\bar{Z}}\right)\\
&={-{\mu\over\pi}T\Tb\over 1-{\mu^2\over\pi^2}T\Tb}\ .
\end{aligned}
\end{equation}
These agree with the map \eqref{TmuTZ} derived from a field-theoretic consideration.


\end{document}